\documentstyle[floats,tighten,preprint,eqsecnum,prd,aps,graphicx]{revtex}

\def\diag{\mathop{\rm diag}}
\def\R#1{{\cal R}[#1]}
\def\tr{\mathop{\rm tr}}
\def\maybebar#1{\smash{\stackrel{\scriptscriptstyle (-)}{#1}}%
\vphantom{\bar{#1}}}

\begin{document}  


\preprint{\parbox[t]{15em}{\raggedleft
FERMILAB-PUB-01/393-T \\
KEK-CP-118 \\
YITP-01-89 \\
hep-lat/0112044\\[2.0em]}}
\draft 

\title{Application of heavy-quark effective theory to lattice QCD:\\ 
II.~Radiative corrections to heavy-light currents}

\author{Junpei Harada,$^1$
Shoji~Hashimoto,$^2$ 
Ken-Ichi~Ishikawa,$^{2,3}$ 
Andreas~S.~Kronfeld,$^{3,4}$ 
Tetsuya~Onogi,$^{1,5}$\cite{TO} and
Norikazu~Yamada$^2$}

\address{$^1$Department of Physics, Hiroshima University,
	Higashi-Hiroshima 739-8526, Japan \\
$^2$High Energy Accelerator Research Organization (KEK),
	Tsukuba 305-0801, Japan \\
$^3$Center for Computational Physics, University of Tsukuba,
	Tsukuba 305-8577, Japan \\
$^4$Theoretical Physics Department,
	Fermi National Accelerator Laboratory, Batavia, Illinois 60510 \\
$^5$Yukawa Institute for Theoretical Physics, Kyoto University, 
	Sakyo-ku, Kyoto 606-8502, Japan}

\date{30 December 2004}
\maketitle

\begin{abstract}
We apply heavy-quark effective theory to separate long- and
short-distance effects of heavy quarks in lattice gauge theory.
In this approach, the inverse heavy-quark mass and the lattice spacing
are treated as short distances, and their effects are lumped into
short-distance coefficients.
We show how to use this formalism to match lattice gauge theory to
continuum QCD, order by order in the heavy-quark expansion.
In this paper, we focus on heavy-light currents.
In particular, we obtain one-loop results for the matching
factors of lattice currents, needed for heavy-quark phenomenology, such
as the calculation of heavy-light decay constants, and heavy-to-light
transition form factors.
Results for the Brodsky-Lepage-Mackenzie scale~$q^*$ are also given.
\end{abstract} 



\section{Introduction}
\label{sec:intro}
A key ingredient in flavor physics is the calculation of hadronic
matrix elements of the electroweak Hamiltonian.
For example, one would like to calculate, from first principles,
quantities such as leptonic decay constants, semi-leptonic form
factors, and the amplitudes for neutral-meson mixing.
Numerical calculations with lattice QCD offer a way to obtain these
quantities, eventually with well-controlled estimates of the numerical
uncertainties~\cite{Kronfeld:1993jf}.

The properties of $B$ and $D$ mesons are especially interesting, but
the relatively large $b$ and $c$ quark masses make it difficult, with
today's computers, to carry out lattice calculations in the limit $m_Q
a\to 0$ for which lattice QCD was first developed.
(Here $m_Q$ is the $b$ or $c$ quark mass,
and $a$ is the lattice spacing.)
One can, however, use the simplifying features of the heavy-quark limit
of QCD to make lattice calculations tractable.
As $m_Q$ is increased far above the typical scale of the wave function,
$\Lambda_{\text{QCD}}$, the hadrons' wave functions depend less and less
on~$m_Q$.
As $m_Q\to\infty$ the wave functions become flavor and spin
symmetric~\cite{Isgur:1989vq}.
For quarkonia similar simplifications occur, including spin
symmetry~\cite{Caswell:1986ui}.

In this paper we construct vector and axial vector currents with
one quark heavy and the other light.
These currents are needed to obtain the decay constants of heavy-light
mesons, and the form factors for decays of the form $H\to Ll\nu_l$,
where $H$ is a charmed or $b$-flavored hadron (\emph{e.g.}, $B$, $D$;
$\Lambda_b$, $\Lambda_c$), decaying to a light hadron~$L$ (\emph{e.g.},
$\pi$, $K$, $\rho$; $p$, etc.)\ and a lepton~$l$ and its
neutrino~$\nu_l$.
In particular, we provide a way to treat radiative and power corrections
consistently.
This paper is a sequel to Ref.~\cite{Kronfeld:2000ck}, which focussed on
power corrections.
Here we discuss the case of heavy-light bilinears in detail, and
we compute explicitly the matching factors for the currents
introduced in the ``Fermilab'' formalism~\cite{El-Khadra:1997mp}.
Heavy-heavy bilinears are considered in a companion 
paper~\cite{Kronfeld:1999tk}.

To interpret lattice calculations when $m_Qa\ll1$, it is convenient to
describe cutoff effects with the Symanzik local effective Lagrangian
(LE${\cal L}$) and expand the LE${\cal L}$'s short-distance coefficients
in powers of $m_Qa$~%
\cite{Symanzik:1979ph,Symanzik:1983dc,Luscher:1996sc,Jansen:1996ck}.
When $m_Qa\not\ll1$, however, one should realize that it is not lattice
gauge theory that breaks down but rather the Symanzik description,
especially its expansion in~$m_Qa$.
If $m_Qa$ is large because $m_Q\gg\Lambda_{\text{QCD}}$, then the
simplifying features of the heavy-quark limit provide an alternative.
Instead of matching lattice gauge theory directly to continuum QCD,
one can match to the heavy-quark effective theory~(HQET) or, for
quarkonia, to non-relativistic QCD (NRQCD).
In this approach, the inverse heavy-quark mass and the lattice spacing
are both treated as short distances, and a simple picture arises, in
which heavy-quark discretization effects are lumped into short-distance
coefficients.
Heavy-quark cutoff effects are systematically reducible, by adjusting
the heavy-quark expansion for lattice gauge theory to agree term-by-term
with continuum~QCD.

Such application of HQET to lattice QCD was started in
Ref.~\cite{Kronfeld:2000ck}, building on Ref.~\cite{El-Khadra:1997mp}.
In this paper we extend the formalism to heavy-light currents.
We use the heavy-quark expansion, as generated by HQET, to derive
matching conditions, which are valid for all $m_Qa$ and to all orders in
the gauge coupling.
Our derivation is explicit for dimension-four currents, which is the
next-to-leading dimension, but generalization to higher-dimension
operators should be clear.

We also present explicit results for the one-loop radiative corrections
to the normalization of the current.
These calculations show that the temporal and spatial components of the
current do not have the same radiative corrections.
This feature has been found
already~\cite{Kuramashi:1998tt,Ishikawa:1997xh},
and the HQET formalism shows why it arises.
In deriving these results we have found a compact way of arranging the
Dirac algebra, which may be useful for calculations with actions, such
as highly improved actions, that are not considered here.

As expected, the coefficients have a strong mass dependence.
Most of this dependence can be handled
non-perturbatively~\cite{Hashimoto:2000yp,Simone:2000nv,El-Khadra:2001rv}.
For equal mass, it is simple to normalize the temporal vector current,
for all masses.
One can then form ratios of renormalization factors, from which the
dominant mass dependence drops out.
Results for these combinations are also presented,
in Sec.~\ref{sec:loop}.

Our one-loop results extend those of Ref.~\cite{Ishikawa:1997xh},
which considered heavy-light currents with the Sheikholeslami-Wohlert
(SW) action~\cite{Sheikholeslami:1985ij} for Wilson
fermions~\cite{Wilson:1975hf} and also with non-relativistic~QCD
(NRQCD).
Results for the Wilson action~\cite{Wilson:1975hf} have been obtained
first by Kuramashi~\cite{Kuramashi:1998tt}.
In Refs.~\cite{Kuramashi:1998tt,Ishikawa:1997xh} a term in the currents,
the so-called rotation
term~\cite{Kronfeld:1995nu,El-Khadra:1997mp}, which is needed for
tree-level improvement at order $1/m_Q$, was omitted.
Here we include the rotation, obtaining the algebraic expression of the
Feynman diagrams for the full Fermilab action.
We present numerical results for the Wilson action (without rotation)
and the SW action (with and without rotation).
These results are appropriate for recent calculations of decay 
constants~\cite{Aoki:1998ji,El-Khadra:1998hq,Bernard:1998xi,%
AliKhan:2000eg,Bernard:2000nv}, which used the radiative corrections 
calculated in Refs.~\cite{Kuramashi:1998tt,Ishikawa:1997xh}.
Our new results have been used in a recent calculation of the form 
factors for the decays $B\to\pi l\nu_l$ and 
$D\to\pi l\nu_l$~\cite{El-Khadra:2001rv}.
We also have obtained results for the Fermilab action on anisotropic
lattices~\cite{Harada:2001ei}.

Our formalism should be useful for computing matching factors (beyond
one-loop) also in lattice~NRQCD~\cite{Lepage:1987gg}.
Applied to the static limit~\cite{Eichten:1987xu}, it generalizes the
formalism of Eichten and Hill~\cite{Eichten:1990zv}.
At one-loop order, similar methods have been developed to calculate the
heavy-light matching coefficients for lattice
NRQCD~\cite{Davies:1993ec,Morningstar:1998ep}.
As in the Symanzik
program~\cite{Symanzik:1979ph,Symanzik:1983dc,Luscher:1996sc}, the
advantage of introducing a continuum effective field theory is that the
formalism provides a clear definition of the matching coefficients at
every order in perturbation theory (in the gauge coupling).
Indeed, it may also provide a foundation for a non-perturbative
improvement program.

This paper is organized as follows.
Section~\ref{sec:match} discusses three ways to separate long and
short distance physics with (continuum) effective field theories.
The first is Symanzik's description of lattice spacing effects;
we also discuss its breakdown when $m_Qa\not\ll1$.
The second is the HQET description of heavy quarks, applied to continuum
QCD.
The third is the HQET description of heavy quarks on the lattice, which
applies when $m_Q\gg\Lambda_{\text{QCD}}$, for all $m_Qa$.
In particular, we obtain a definition of the matching factors for
the vector and axial-vector heavy-light currents.
Section~\ref{sec:match} also shows how the HQET matching procedure
is related to the Symanzik procedure in the regime where both apply.
Then, the Fermilab action is reviewed in Sec.~\ref{sec:lattice},
and in Sec.~\ref{sec:loop} we present one-loop results for the
matching factors.
Some concluding remarks are made in Sec.~\ref{sec:conclusions}.
Three appendices contain details of the one-loop calculation,
including an outline of a method to obtain compact expressions,
and explicit results for the one-loop Feynman integrands for the
renormalization factors with the full Fermilab action.

Instead of printing tables of the numerical results in
Sec.~\ref{sec:loop}, we are making a suite of programs freely
available\cite{p:epaps}.
This suite includes programs for the heavy-heavy currents treated in our
companion paper~\cite{Kronfeld:1999tk}.

\section{Matching to Continuum Field Theories}
\label{sec:match}

In this section we discuss how to interpret the physical content of
lattice field theories by matching to continuum field theories.
First, the standard Symanzik formalism for describing cutoff effects
is reviewed, and we recall how this description breaks down for
heavy quarks.
After reviewing the HQET description of (continuum)~QCD, we adapt HQET
to describe lattice gauge theory.
Comparison of the two then yields a matching procedure that is valid
whenever $m_Q\gg\Lambda_{\text{QCD}}$, and for all~$m_Qa$.
In the limit $m_Qa\ll1$ both the HQET and the Symanzik descriptions
should apply, so we are able to derive relations between the some of
the matching coefficients.

\subsection{Symanzik Formalism}
\label{subsec:sym}

The customary way to define matching factors for lattice gauge theory
is to apply Symanzik's formalism.
Then the short-distance lattice artifacts are described by a local
effective Lagrangian (LE${\cal L}$) and local effective operators.
For the Lagrangian of any lattice field theory one can
write~\cite{Symanzik:1979ph,Symanzik:1983dc}
\begin{equation}
	{\cal L}_{\text{lat}} \doteq {\cal L}_{\text{Sym}},
	\label{eq:lat=Sym}
\end{equation}
where the symbol $\doteq$ can be read
``has the same on-shell matrix elements as''.
The left-hand side is a lattice field theory, and the right-hand side is
a continuum field theory, whose ultraviolet behavior is regulated and
renormalized completely separately from the lattice of the left-hand
side.
The LE${\cal L}$ is the Lagrangian of the corresponding continuum
field theory, plus extra terms to describe discretization effects.
For lattice QCD
\begin{equation}
	{\cal L}_{\text{Sym}} = {\cal L}_{\text{QCD}} + {\cal L}_I,
		\label{eq:LEL}
\end{equation}
where ${\cal L}_{\text{QCD}}$
is the renormalized, continuum QCD Lagrangian.
We focus on the quarks, so for our purposes
\begin{equation}
	{\cal L}_{\text{QCD}} = - \bar{q}\left(
		{\kern+0.1em /\kern-0.65em D} + m_q\right)q.
	\label{eq:QCD}
\end{equation}
Lattice artifacts are described by higher-dimension operators, 
\begin{equation}
	{\cal L}_I = aK_{\sigma\cdot F} \bar{q} i\sigma_{\mu\nu}F^{\mu\nu} q
		+ \cdots, \label{eq:artifacts}
\end{equation}
where $a$ is the lattice spacing and $K_{\sigma\cdot F}$ is a
short-distance coefficient that depends on details of the lattice
action~\cite{Luscher:1996sc}.
The lattice artifacts in ${\cal L}_I$ can be treated as a
perturbation.
In this way a series can be developed, with matrix elements in the
(continuum) eigenstates of~${\cal L}_{\text{QCD}}$.
Equation~(\ref{eq:LEL}) omits dimension-five operators of the form
$\bar{q}R({\kern+0.1em /\kern-0.65em D} + m_q)q$ or
$\bar{q}(-\loarrow{\kern+0.1em /\kern-0.65em D} + m_q)Rq$, for
arbitrary~$R$, which make no contribution to on-shell matrix elements,
owing to the equations of motion implied by Eq.~(\ref{eq:QCD}).

The vector and axial vector currents can be described in a similar way.
Consider, for example, the flavor-changing transition $s\to u$.
Then one may write~\cite{Luscher:1996sc}
\begin{eqnarray}
	V^\mu_{\text{lat}} & \doteq & Z_V^{-1} {\cal V}^\mu -
		aK_V \partial_\nu \bar{u} \sigma^{\mu\nu}s + \cdots ,
	\label{eq:LEV} \\
	A^\mu_{\text{lat}} & \doteq & Z_A^{-1} {\cal A}^\mu +
		aK_A \partial^\mu \bar{u}i          \gamma_5 s + \cdots ,
	\label{eq:LEA}
\end{eqnarray}
where
\begin{eqnarray}
	{\cal V}^\mu & \equiv & \bar{u}i\gamma^\mu     s ,
	\label{eq:V} \\
	{\cal A}^\mu & \equiv & \bar{u}i\gamma^\mu\gamma_5 s ,
	\label{eq:A}
\end{eqnarray}
are the vector and axial vector currents in QCD.
Further dimension-four operators are omitted, because they are linear
combinations of those listed and others that vanish by the equations of
motion.
Like the terms of dimension five and higher in ${\cal L}_I$, the
dimension-four currents can be treated as perturbations.
Matrix elements of $Z_VV^\mu_{\text{lat}}$ and $Z_AA^\mu_{\text{lat}}$
then give those of continuum QCD, at least in the limit $a\to 0$.

The short-distance coefficients---%
$K_{\sigma\cdot F}$, $K_J$, and $Z_J$ ($J=V$, $A$)---%
are, in general, functions of the gauge coupling and the quark masses
(in lattice units), and they depend on the renormalization scheme of
the~LE$\cal L$.
For $m_qa\ll1$ ($q=u$, $d$, $s$), it is consistent and satisfactory to
replace~$K_{\sigma\cdot F}$ and~$K_J$ with their values at $m_qa=0$, and
to replace the~$Z_J$ with the first two terms of the Taylor expansion
around $m_qa=0$.
For example, with Wilson
fermions~\cite{Wilson:1975hf,Sheikholeslami:1985ij} and conventional
bilinears for the lattice currents, one finds $K_V^{[0]}=K_A^{[0]}=0$,
and
\begin{eqnarray}
	K_{\sigma\cdot F}^{[0]} & = & \case{1}{4}(1-c_{\text{SW}})
		+ O(ma), \label{eq:Ctree} \\
	Z_V^{[0]}  =  Z_A^{[0]} & = & 1 + \case{1}{2}(m_u+m_s)a
		+ O(m^2a^2), \label{eq:Ztree}
\end{eqnarray}
where the superscript~``$[0]$'' denotes the tree level,
and $c_{\text{SW}}$ is the clover coupling of the SW
action~\cite{Sheikholeslami:1985ij} (cf.\ Sec.~\ref{sec:lattice}).
Moreover, in the hands of the
{\sl Alpha} Collaboration~\cite{Luscher:1996sc,Jansen:1996ck},
Eqs.~(\ref{eq:lat=Sym})--(\ref{eq:LEA}) are the foundation of a
non-perturbative procedure for adjusting $K_{\sigma\cdot F}$, $K_V$,
and~$K_A$ to be of order~$aM_p$, where $M_p$ is a (light) hadronic
mass scale, and also for computing $Z_V$ and $Z_A$ non-perturbatively
(through order~$M_pa$).
Then all lattice artifacts in the mass spectrum, decay constants, and
form factors are of order~$a^2$.

For a heavy flavor~$Q$, however, it is not practical to keep $m_Qa$
small enough so that this program straightforwardly applies.
Recent work that uses the fully $O(a)$-improved action and currents has
chosen the heavy-quark mass $m_Qa$ to be as large as 0.7 or~so.
Thus, $(m_Qa)^2$ is not small,%
\footnote{Also, the lowest chosen values of $m_Q$ are around
1~GeV, which may be too small to be considered ``heavy''.}
and one should check whether contributions of order~$(m_Qa)^2$ are under
control.
Indeed, if one keeps the full mass dependence in the coefficients, one
finds that the simple description of Eqs.~(\ref{eq:LEL})--(\ref{eq:LEA})
breaks down.
The relation between energy and momentum
becomes~\cite{El-Khadra:1997mp,Mertens:1998wx}
\begin{equation}
	E^2(\bbox{p}) = m_1^2 + \frac{m_1}{m_2}\bbox{p}^2 + O(p^4a^2),
	\label{eq:Emp}
\end{equation}
where, for the Wilson and SW actions,
\begin{eqnarray}
	m^{[0]}_1a & = & \ln(1+m_0a), \label{eq:m1} \\
	\frac{1}{m^{[0]}_2a} & = &
		\frac{2}{m_0a(2+m_0a)} + \frac{1}{1+m_0a}, \label{eq:m2Wilson}
\end{eqnarray}
and $m_0$ is the bare lattice mass.
Generalizations valid at every order in perturbation theory also have
been derived~\cite{Mertens:1998wx}.
In a similar vein, the spatial and temporal components of the currents
no longer take the same matching coefficients, as shown by explicit
one-loop calculations~\cite{Kuramashi:1998tt,Ishikawa:1997xh}.

The energy-momentum relation in Eq.~(\ref{eq:Emp}) is obtained for
$\bbox{p}a\ll1$ but $m_Qa\not\ll1$.
It can be described by modifying the standard LE${\cal L}$ to
\begin{equation}
	{\cal L}_{\text{lat}} \doteq - \bar{q}\left(\gamma_4 D_4 +
		\sqrt{\frac{m_1}{m_2}} \bbox{\gamma}\cdot\bbox{D} + m_1\right)q +
		{\cal L}'_I,
	\label{eq:LELmod}
\end{equation}
that is, temporal and spatial directions must be treated asymmetrically
in the dimension-four Lagrangian, and also in the higher-dimension
terms~${\cal L}'_I$.
From the tree-level formulas, Eqs.~(\ref{eq:m1}) and~(\ref{eq:m2Wilson}),
\begin{equation}
	\frac{m_1}{m_2} = 1 - \case{2}{3}m_1^2a^2 +
		\case{1}{2}m_1^3a^3 + \cdots,
	\label{eq:m1m2}
\end{equation}
so one sees that the deviation from the standard description is of
order~$(ma)^2$.
(At the one-loop order~\cite{Mertens:1998wx}, and at every order
in~$g^2$, Eq.~(\ref{eq:m1m2}) still has no term linear in~$ma$.)
One can arrive at Eqs.~(\ref{eq:LELmod}) and~(\ref{eq:m1m2}) also by
starting with Eq.~(\ref{eq:LEL}), including higher-dimension terms,
and eliminating~$\gamma_4D_4^3$ and~$D_4^4$, etc., by applying the
equations of motion.

In any case, deviations of $m_1/m_2$---and similar ratios---from~1 are
present in lattice calculations.
With the Wilson or SW actions $1-m_1/m_2$, for example, is 10~percent or
greater for $m_1a>0.6$.
Although this numerical estimate is made at the tree level, it is
implausible that radiative corrections or bound-state effects could
wash the error away.
In summary, the description of Eqs.~(\ref{eq:LEL})--(\ref{eq:LEA})
is no longer accurate when $m_Qa\not\ll1$.

There are several possible remedies.
One is to do numerical calculations with $a$ so small that,
even for the $b$ quark, $m_ba\ll1$.
Despite the exponential growth in computer power, this remedy will not
be available for many years.
Another remedy is to add a parameter to the lattice action, which can be
tuned to set $m_1=m_2$~\cite{El-Khadra:1997mp}.
An example of this is an action with two hopping parameters.
Then, the continuum description can again take the form in
Eq.~(\ref{eq:LEL}), starting with the continuum~${\cal L}_{\text{QCD}}$,
although it is still useful to describe the higher-dimension terms
asymmetrically.
A third remedy is to realize that it is the \emph{description}, rather
than the underlying lattice gauge theory, that has broken down.
Since lattice gauge theory with Wilson fermions has a well-behaved
heavy-quark limit~\cite{El-Khadra:1997mp}, it is possible to use
heavy-quark effective theory (HQET) or NRQCD to describe short-distance
effects, including the lattice artifacts of the heavy
quark~\cite{Kronfeld:2000ck}.
This last remedy is explained in detail in Sec.~\ref{subsec:hqetlat},
where we show also how all three strategies are connected.

\subsection{HQET description of QCD}
\label{subsec:hqetcont}

The breakdown of the standard Symanzik description of cutoff effects for
Wilson fermions arises because the kinematics of heavy hadron decays
single out a vector, namely, the heavy hadron velocity.
But, since the heavy-quark mass is also much larger than the spatial
momenta of the problem, the dynamics simplify.
In continuum QCD, this has led to the development of the 
effective field theories
HQET~\cite{Eichten:1987xu,Eichten:1990zv,Grinstein:1990mj,Georgi:1990um%
,Eichten:1990vp}
and NRQCD~\cite{Caswell:1986ui,Lepage:1987gg}.
These two effective theories are useful for generating an expansion in
$\bbox{p}/m_Q$.
They share a common effective Lagrangian, but the power in
$\bbox{p}/m_Q$ assigned to any given operator is not necessarily
the same.
In HQET the power can be deduced immediately from the dimension,
whereas in NRQCD it is deduced by counting powers of the relative
velocity of the $\bar{Q}Q$ system.
The discussion in this paper will follow the counting of~HQET,
but the logic could be repeated with the counting of~NRQCD.

Our aim is to show, for the case of heavy-light
currents, how to use HQET to extend the standard Symanzik program into
the region where $m_Qa$ is no longer small.
This program was started in Ref.~\cite{Kronfeld:2000ck}, building on
Ref.~\cite{El-Khadra:1997mp}.
The formalism holds for all~$m_Qa$, but, like the usual HQET,
it requires
\begin{equation}
	m_Q \gg \bbox{p},\, \Lambda_{\text{QCD}}.
	\label{eq:when}
\end{equation}
First, in this subsection, we recall the HQET description of continuum
QCD, paralleling the discussion in Sec.~\ref{subsec:sym}.
Then, in Sec.~\ref{subsec:hqetlat}, we explain what changes are needed
to describe the cutoff effects of lattice NRQCD and of lattice gauge
theory with Wilson fermions.

The HQET conventions are the same as those given Sec.~III of
Ref.~\cite{Kronfeld:2000ck}.
The velocity needed to construct HQET is~$v$.
The fourth Euclidean component $v_4=iv^0$, so in the rest frame
$v=(i,\bbox{0})$.
The metric is taken to be $\diag(\pm 1,1,1,1)$, with the upper (lower)
sign for Euclidean (Minkowski) spacetime.
In either case, $v^2=-1$.
The heavy quark field is called $h_v$, and it satisfies the constraint
$\case{1}{2}(1-i\kern+0.1em /\kern-0.55em v)h_v = h_v$, or
\begin{equation}
	         \kern+0.1em /\kern-0.55em v h_v = ih_v, \quad
	\bar{h}_v\kern+0.1em /\kern-0.55em v     = i\bar{h}_v.
	\label{eq:Pvh}
\end{equation}
Physically Eq.~(\ref{eq:Pvh}) means that $h_v$ describes only quarks,
but not anti-quarks.
The tensor $\eta^\mu_\nu=\delta^\mu_\nu+v^\mu v_\nu$ projects onto
components orthogonal to~$v$.
For a vector~$p$, the component orthogonal to~$v$ is
$p_\perp^\mu=\eta^\mu_\nu p^\nu=p^\mu+v^\mu{(v\cdot p)}$;
in the rest frame, these are the spatial components.

HQET describes the dynamics of heavy-light bound states with an
effective Lagrangian built from~$h_v$.
So, for these states, one can say
\begin{equation}
	{\cal L}_{\text{QCD}} \doteq {\cal L}_{\text{HQET}},
	\label{eq:QCD=HQET}
\end{equation}
where
\begin{equation}
	{\cal L}_{\text{HQET}} =
	{\cal L}^{(0)} +
	{\cal L}^{(1)} +
	{\cal L}^{(2)} + \cdots.
	\label{eq:L}
\end{equation}
For HQET ${\cal L}^{(s)}$ contains terms of dimension $4+s$.
Note that the ultraviolet regulator and renormalization scheme of the
two sides of Eq.~(\ref{eq:QCD=HQET}) need not be the same, although
dimensional regularization and the $\overline{\rm MS}$ scheme are
usually used for both.

For this paper it is enough to consider the first two
terms, ${\cal L}^{(0)}$ and ${\cal L}^{(1)}$.
The leading, dimension-four term is 
\begin{equation}
	{\cal L}^{(0)} = \bar{h}_v(iv\cdot D - m)h_v.
	\label{eq:L0}
\end{equation}
The choice of $v$ is somewhat arbitrary.
If $v$ is close to the heavy quark's velocity,%
\footnote{In NRQCD applications the relative velocity between the heavy
quark and heavy anti-quark should not be confused with the velocity~$v$
introduced here.}
then ${\cal L}^{(0)}$ is a good starting point for the heavy-quark
expansion, which treats the higher-dimension operators as small.
The most practical choice is the containing hadron's velocity.

The mass term in ${\cal L}^{(0)}$ is often omitted.
By heavy-quark symmetry, it has an effect neither on bound-state wave
functions nor, consequently, on matrix elements.
It does affect the mass spectrum, but only additively.
Including the mass obscures the heavy-quark flavor symmetry,
but only slightly~\cite{Kronfeld:2000ck}.
When the mass term is included, higher-dimension operators are
constructed with ${\cal D}^\mu=D^\mu-imv^\mu$~\cite{Falk:1992fm}.
To describe on-shell matrix elements one may omit operators that 
vanish by the equation of motion, $-iv\cdot{\cal D}h_v=0$, derived 
from Eq.~(\ref{eq:L0}).
In practice, therefore, higher-dimension operators are constructed 
from ${\cal D}^\mu_\perp=D^\mu_\perp$ and
$[{\cal D}^\mu,{\cal D}^\nu]=[D^\mu,D^\nu]=F^{\mu\nu}$.

The dimension-five interactions are
\begin{equation}
	{\cal L}^{(1)} = {\cal C}_2 {\cal O}_2 +
		{\cal C}_{\cal B} {\cal O}_{\cal B},
	\label{eq:L1}
\end{equation}
where ${\cal C}_2$ and ${\cal C}_{\cal B}$ are short-distance
coefficients, and
\begin{eqnarray}
	{\cal O}_2 & = &
		\bar{h}_vD_\perp^2 h_v, \label{eq:O2} \\
	{\cal O}_{\cal B} & = & 
		\bar{h}_v s_{\alpha\beta}B^{\alpha\beta}h_v,
		\label{eq:OB}
\end{eqnarray}
with $s_{\alpha\beta}=-i\sigma_{\alpha\beta}/2$
and  $B^{\alpha\beta}=\eta^\alpha_\mu\eta^\beta_\nu F^{\mu\nu}$.

In Eq.~(\ref{eq:L0}) one should think of the quark mass $m$ as 
a short-distance coefficient.
By reparametrization invariance~\cite{Luke:1992cs}, the same mass
appears in the denominator of the kinetic energy~${\cal C}_2{\cal O}_2$,
namely,
\begin{equation}
	{\cal C}_2 = \frac{1}{2m}.
\end{equation}
If operator insertions in HQET are renormalized with a minimal 
subtraction in dimensional regularization, then $m$ is the 
(perturbative) pole mass.
With other ultraviolet regulators, the operator and the mass~$m$
could become $\mu$-dependent.
Even in mass-independent schemes, the chromomagnetic
operator~${\cal O}_{\cal B}$ depends on the renormalization point~$\mu$
of the HQET, and that dependence is canceled by
\begin{equation}
	{\cal C}_{\cal B}(\mu) = \frac{z_{\cal B}(\mu)}{2m},
\end{equation}
with $2m$ appearing so that $z_{\cal B}$ is unity at the tree level.

The description of electroweak flavor-changing operators proceeds along
the same lines.
The flavor-changing vector current for a $b\to q$ transition, defined
to be ${\cal V}^\mu=\bar{q}i\gamma^\mu b$ as in Eq.~(\ref{eq:V}),
is described in HQET by
\begin{equation}
	{\cal V}^\mu \doteq C_{V_\parallel} v^\mu \bar{q}h_v  +
		C_{V_\perp} \bar{q}i\gamma^\mu_\perp h_v -
		\sum_{i=1}^6 B_{Vi} {\cal Q}^\mu_{Vi} + \cdots,
	\label{eq:VcontHQET}
\end{equation}
where $h_v$ is the HQET field, which satisfies Eq.~(\ref{eq:L}) and
whose dynamics are given by~${\cal L}_{\text{HQET}}$.
The dimension-four operators are
\begin{eqnarray}
	{\cal Q}^\mu_{V1} & = & - v^\mu \bar{q}
		{\kern+0.1em /\kern-0.65em D}_\perp h_v, \label{eq:Q1} \\
	{\cal Q}^\mu_{V2} & = & \bar{q}i\gamma^\mu_\perp
		{\kern+0.1em /\kern-0.65em D}_\perp h_v, \label{eq:Q2} \\
	{\cal Q}^\mu_{V3} & = & \bar{q}iD^\mu_\perp h_v, \label{eq:Q3} \\
	{\cal Q}^\mu_{V4} & = & + v^\mu \bar{q}
		{\kern+0.1em /\kern-0.65em \loarrow{D}}_\perp h_v, \label{eq:Q4} \\
	{\cal Q}^\mu_{V5} & = & \bar{q}
		{\kern+0.1em /\kern-0.65em \loarrow{D}}_\perp
		i\gamma^\mu_\perp h_v,  \label{eq:Q5} \\
	{\cal Q}^\mu_{V6} & = & \bar{q}i\loarrow{D}^\mu_\perp h_v. \label{eq:Q6}
\end{eqnarray}
Further dimension-four operators are again omitted, because they
are linear combinations of those listed and others that vanish by the
equations of motion.
For example,
$\bar{q}(iv\cdot\loarrow{D}) v^\mu h_v =
\bar{q}({\kern+0.1em /\kern-0.65em \loarrow{D}}_\perp -
{\kern+0.1em /\kern-0.65em \loarrow{D}})v^\mu h_v =
{\cal Q}^\mu_{V4}-m_qv^\mu\bar{q}h_v$,
where the Dirac equation is used for the last step.

The axial vector current ${\cal A}^\mu=\bar{q}i\gamma^\mu\gamma_5b$ has
a completely analogous description,
\begin{equation}
	{\cal A}^\mu \doteq
		C_{A_\perp} \bar{q}i\gamma^\mu_\perp \gamma_5h_v -
		C_{A_\parallel} v^\mu \bar{q}\gamma_5h_v  -
		\sum_{i=1}^6 B_{Ai} {\cal Q}_{Ai}^\mu + \cdots,
	\label{eq:AcontHQET}
\end{equation}
where each operator ${\cal Q}_{Ai}^{\mu}$ is obtained from
${\cal Q}_{Vi}^{\mu}$ by replacing $\bar{q}$ with~$-\bar{q}\gamma_5$.

The short-distance coefficients of HQET depend on the heavy-quark
mass~$m$, as well as $\mu/m$ and $m_q/m$, where $\mu$ is the the
renormalization scale and $m_q$ is the light quark mass.
They are not explicitly needed in this paper, but it may be instructive
to give the coefficients of the dimension-three terms through one-loop
order, with $m_q=0$ ($J=V$, $A$)~\cite{Eichten:1990zv}:
\begin{eqnarray}
	C_{J_\parallel} & = & 1 + \frac{g^2C_F}{16\pi^2}\left(
		\gamma_h\ln(m^2/\mu^2) - 2 \right),
	\label{eq:Cparallel} \\
	C_{J_\perp}     & = & 1 + \frac{g^2C_F}{16\pi^2}\left(
		\gamma_h\ln(m^2/\mu^2) - 4 \right),
	\label{eq:Cperp}
\end{eqnarray}
where the anomalous dimension $\gamma_h=3/2$.
The $\mu$-independent part of $C_{A_\parallel}$ and $C_{A_\perp}$
given here assumes that the axial current is renormalized in a chirally
symmetric way~\cite{Trueman:1995ca}.
The coefficients of the dimension-four currents are
\begin{eqnarray}
	B_{Ji}^{[0]} & = & \frac{1}{2m},\quad i\le 2, \label{eq:B1B2} \\
	B_{Ji}^{[0]} & = & 0,           \quad i\ge 3, \label{eq:B3-6}
\end{eqnarray}
at the tree level, but all $B_{Ji}$ become non-trivial when radiative
corrections are included.

\subsection{HQET description of lattice gauge theory}
\label{subsec:hqetlat}

HQET provides a systematic way to separate the short distance~$1/m$
from the scale $\Lambda_{\text{QCD}}$ in heavy-light matrix elements,
as long as the condition~(\ref{eq:when}) holds.
The formalism can also be applied to lattice gauge theory, again as long
as condition~(\ref{eq:when}) holds (and $\bbox{p}a\ll1$).
When lattice NRQCD is used for heavy-light systems, this is because
${\cal L}_{\text{HQET}}$ is just the Symanzik LE${\cal L}$ for lattice
NRQCD.
When Wilson fermions are used for heavy quarks, one may also apply HQET,
because they have the same particle content and heavy-quark
symmetries~\cite{Kronfeld:2000ck}.
In both cases bilinears of lattice fermions fields are introduced to
approximate the continuum QCD currents.
One field corresponds to the light quark, and the other to the
heavy quark.
An explicit construction, through order $1/m$, is in
Ref.~\cite{Morningstar:1998ep} for lattice NRQCD, and a similar
construction for Wilson fermions is in Sec.~\ref{sec:lattice}.
Lattice artifacts stemming from the light quark can be described as in
Sec.~\ref{subsec:sym}, but lattice artifacts of the heavy quark should be
lumped into the HQET short-distance coefficients.
Some of the operators needed to describe heavy-quark
discretization effects do not appear in the usual HQET description
of continuum~QCD.
For example, the dimension-seven operator $\sum_i\bar{h}_vD_i^4h_v$
(written here in the rest frame) appears in~${\cal L}^{(3)}$ to
describe the breaking of rotational invariance on the lattice.
Similarly, at and beyond dimension five there are HQET current
operators to describe violations of rotational symmetry in the
lattice currents.
Because of the high dimension, these effects lie beyond the scope
of this paper, which concentrates on operators of leading and
next-to-leading dimension.

In this way, the preceding description of continuum QCD can be repeated
for lattice gauge theory with the same logic and structure.
Instead of Eq.~(\ref{eq:lat=Sym}), one introduces a relation like
Eq.~(\ref{eq:QCD=HQET}),
\begin{equation}
	{\cal L}_{\text{lat}} \doteq {\cal L}_{\text{HQET}},
\end{equation}
where ${\cal L}_{\text{lat}}$ is a lattice Lagrangian for NRQCD or
Wilson quarks, and ${\cal L}_{\text{HQET}}$ is an HQET Lagrangian
with the same operators as in Eqs.~(\ref{eq:L0}) and (\ref{eq:L1}),
but modified coefficients.
In the dimension-four HQET Lagrangian ${\cal L}^{(0)}$, one must now
replace $m$ with the heavy quark rest mass~$m_1$.
The other coefficients will be denoted ${\cal C}_i^{\text{lat}}$.
In particular, in ${\cal L}^{(1)}$ the coefficient of the kinetic energy
becomes
\begin{equation}
	{\cal C}_2^{\text{lat}} = \frac{1}{2m_2}.
\end{equation}
If operator insertions of ${\cal O}_2$ continue to be defined by
dimensional regularization with minimal subtraction, then both the
rest mass~$m_1$ and the kinetic mass~$m_2$ generalize the perturbative
pole mass.
Like the usual pole mass, they are properties of the pole in the
perturbative quark propagator~\cite{Mertens:1998wx}, and they are
infrared finite and gauge independent~\cite{Kronfeld:1998di}.
The lattice breaks Lorentz (or Euclidean) invariance, so
reparametrization invariance no longer requires $m_2$ to be the same
as~$m_1$.

Similarly, a heavy-light lattice (axial) vector current
$V^\mu_{\text{lat}}$ ($A^\mu_{\text{lat}}$) can be described by
\begin{eqnarray}
	V^\mu_{\text{lat}} & \doteq &
		C_{V_\parallel}^{\text{lat}} v^\mu \bar{q}h_v  +
		C_{V_\perp}^{\text{lat}} \bar{q}i\gamma^\mu_\perp h_v -
		\sum_{i=1}^6 B_{Vi}^{\text{lat}} {\cal Q}^\mu_{Vi} + \cdots,
	\label{eq:VlatHQET} \\
	A^\mu_{\text{lat}} & \doteq &
		C_{A_\perp}^{\text{lat}} \bar{q}i\gamma^\mu_\perp\gamma_5h_v -
		C_{A_\parallel}^{\text{lat}} v^\mu \bar{q}\gamma_5h_v  -
		\sum_{i=1}^6 B_{Ai}^{\text{lat}} {\cal Q}^\mu_{Ai} + \cdots,
	\label{eq:AlatHQET}
\end{eqnarray}
but there are two important changes from Eq.~(\ref{eq:VcontHQET}).
First, the light quarks (and gluons) are now also on the lattice,
so they are described by their usual Symanzik LE${\cal L}$s.
Second, the short-distance coefficients of HQET are modified,
because the lattice modifies the dynamics at short distances.
The coefficients $C_{J_\parallel}^{\text{lat}}$, 
$C_{J_\perp}^{\text{lat}}$, and $B_{Ji}^{\text{lat}}$ now depend on 
the lattice spacing~$a$, \emph{i.e.}, on~$ma$, in addition to $m$, 
$\mu/m$, and $m_q/m$.
A~heavy-light lattice axial vector current has an analogous description.

On the other hand, in Eqs.~(\ref{eq:VcontHQET}) and~(\ref{eq:VlatHQET})
the HQET operators are the same.
As a rule, the ultraviolet regulator of an effective theory does not
have to be the same as that of the underlying theory.
(The standard Symanzik program works this way.)
Thus, when describing lattice gauge theory one is free to regulate
HQET just as one would when describing continuum QCD.
Moreover, since Eqs.~(\ref{eq:Q1})--(\ref{eq:Q6}) give a complete set of
dimension-four HQET currents, the coefficients
$C_{J_\parallel,J_\perp}^{\text{lat}}$ and $B_{Ji}^{\text{lat}}$ contain
short-distance effects from both the light and the heavy sectors.

By comparing the HQET descriptions of lattice and continuum QCD,
one can see how lattice matrix elements differ from their continuum
counterparts.
The continuum matrix element of $v\cdot{\cal V}$, for example, is
\begin{eqnarray}
	\langle L|v\cdot {\cal V}|B\rangle & = & -
		C_{V_\parallel} \langle L|\bar{q}h_v|B_v^{(0)}\rangle
			- 
		B_{V1} \langle L|v\cdot {\cal Q}_{V1}|B_v^{(0)}\rangle
			- 
		B_{V4} \langle L|v\cdot {\cal Q}_{V4}|B_v^{(0)}\rangle
			\nonumber \\ & - &
		{\cal C}_2        C_{V_\parallel} \int d^4x
			\langle L|T\,{\cal O}_2(x) \bar{q}h_v|B_v^{(0)}\rangle^\star
			- 
		{\cal C}_{\cal B} C_{V_\parallel} \int d^4x
			\langle L|T\,{\cal O}_{\cal B}(x)
			\bar{q}h_v|B_v^{(0)}\rangle^\star 
			\nonumber \\ & + &
		O(\Lambda^2/m^2), \label{eq:VHQE}
\end{eqnarray}
where $L$ is any light hadronic state, including the vacuum.
(The $\star$-ed $T$~product is defined in Ref.~\cite{Kronfeld:2000ck};
this detail is unimportant here.)
On the left-hand  side $B$         denotes a $b$-flavored hadron, and
on the right-hand side $B_v^{(0)}$ denotes the corresponding eigenstate
of the leading effective Lagrangian~${\cal L}^{(0)}$.
Similarly, the lattice matrix element is~\cite{Kronfeld:2000ck}
\begin{eqnarray}
	\langle L|v\cdot V_{\text{lat}}|B\rangle & = & -
		C^{\text{lat}}_{V_\parallel} \langle L|\bar{q}h_v|B_v^{(0)}\rangle
			- 
		B^{\text{lat}}_{V1} \langle L|v\cdot {\cal Q}_{V1}|B_v^{(0)}\rangle
			- 
		B^{\text{lat}}_{V4} \langle L|v\cdot {\cal Q}_{V4}|B_v^{(0)}\rangle
			\nonumber \\ & - &
		{\cal C}^{\text{lat}}_2        C^{\text{lat}}_{V_\parallel} \int d^4x
			\langle L|T\,{\cal O}_2(x) \bar{q}h_v|B_v^{(0)}\rangle^\star
			- 
		{\cal C}^{\text{lat}}_{\cal B} C^{\text{lat}}_{V_\parallel} \int d^4x
			\langle L|T\,{\cal O}_{\cal B}(x)
			\bar{q}h_v|B_v^{(0)}\rangle^\star 
			\nonumber \\ & - &
		K_{\sigma\cdot F} C^{\text{lat}}_{V_\parallel} \int\! d^4x
			\langle L|T\,\bar{q}i\sigma Fq(x)
			\bar{q}h_v|B_v^{(0)}\rangle^\star 
			+ 
		O(\Lambda^2a^2b(ma)).
\end{eqnarray}
Compared to Eq.~(\ref{eq:VHQE}), the short-distance coefficients are
modified to depend on~$ma$, there is an extra term from the Symanzik
LE${\cal L}$ of the light quark, and the next power corrections can,
in general, be multiplied by a (bounded) function of~$ma$.
The matrix elements on the right-hand sides are, however, identical,
because in both cases they are defined with
${\cal L}^{(0)}$ describing the heavy quark and
${\cal L}_{\text{QCD}}$ describing the light quark (and gluons).

Similar equations hold for matrix elements of~${\cal V}_\perp$
and~${V_\perp}_{\text{lat}}$, and for the axial vector current.
If one multiplies the equations for the lattice matrix elements with
\begin{eqnarray}
	Z_{J_\parallel} & = &
		\frac{C_{J_\parallel}}{C_{J_\parallel}^{\text{lat}}},
	\label{eq:defZVpara} \\
	Z_{J_\perp}     & = &
		\frac{C_{J_\perp}}{C_{J_\perp}^{\text{lat}}},
	\label{eq:defZVperp}
\end{eqnarray}
and subtracts the result from the continuum equations, one finds
that the difference can be traced solely to the mismatch of the
short-distance coefficients, or
\begin{eqnarray}
	\delta{\cal C}_i = {\cal C}_i^{\text{lat}} & - & {\cal C}_i,
	\label{eq:deltaC} \\
	\delta B_{Ji} =
		Z_{Ji}  B_{Ji}^{\text{lat}} & - & B_{Ji},
	\label{eq:deltaB}
\end{eqnarray}
where the normalization factors $Z_{Ji}$ are
$Z_{J_\parallel}$ for $i=1$,~4, and
$Z_{J_\perp}$     for $i=2$, 3, 5,~6.
In Eqs.~(\ref{eq:deltaC}) and~(\ref{eq:deltaB}) a picture emerges,
where \emph{heavy-quark lattice artifacts are isolated}
into~$\delta{\cal C}_i$ and~$\delta B_{Ji}$.
Furthermore, the analysis presented here makes no explicit reference to
any method for computing the short-distance coefficients, so it applies
at every order in perturbation theory (in $g^2$) and, presumably, at a
non-perturbative level as well.

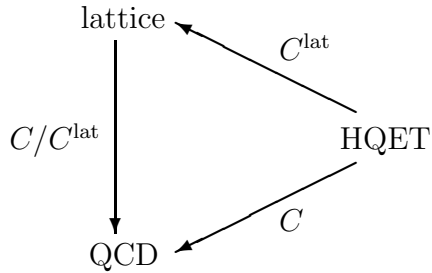
\begin{figure}[!b]
	\centering
	\begin{picture}(200,120)(50,0)
		\thicklines
		\put(102,100){lattice}
		\put(115,95){\vector(0,-1){73}}
		\put(206,68){\vector(-2,1){68}}
		\put(200,54){HQET}
		\put(206,49){\vector(-2,-1){68}}
		\put(105,10){QCD}
		\put(75,54){$C/C^{\text{lat}}$}
		\put(177,88){$C^{\text{lat}}$}
		\put(177,23){$C$}
	\end{picture}
	\caption[fig:matching]{Diagram illustrating how the matching
	factors $C^{\text{lat}}$, $C$, and $Z=C/C^{\text{lat}}$ match
	lattice gauge theory and QCD to HQET, and to each other.}
	\label{fig:matching}
\end{figure}

The matching factors $Z_{J_\parallel}$ and~$Z_{J_\perp}$ play the
following role, sketched in Fig.~\ref{fig:matching}.
In each case, the denominator converts a lattice-regulated scheme to a
renormalized HQET scheme, and the numerator converts the latter to a
renormalized (continuum) QCD scheme.
As long as the same HQET scheme is used, HQET drops out of the
calculation of $Z_{J_\parallel}$ and~$Z_{J_\perp}$.
Moreover, changes in continuum renormalization conventions modify only
the numerator, and changes in the lattice action or currents modify only
the denominator.
In a similar way, dependence on the HQET renormalization scheme drops out
when computing~$\delta{\cal C}_i$ and~$\delta B_{Ji}$.

One can derive a connection between the matching coefficients of the
HQET and the Symanzik descriptions when $ma\ll 1$ and $m\gg\bbox{p}$,
so that both formalisms apply.
With the Lagrangian, one applies HQET to
Eqs.~(\ref{eq:LEL})--(\ref{eq:artifacts}) and identifies the
short-distance coefficients with $m_1$, ${\cal C}_2^{\text{lat}}$,
and~${\cal C}_{\cal B}^{\text{lat}}$.
Then one finds,
\begin{eqnarray}
	m_{1b} & = & m_b + O(a^2), \label{eq:m1=m}  \\
	m_{2b} & = & m_b + O(a^2), \label{eq:m2=m}  \\
	z_{\cal B}^{\text{lat}} & = &
		z_{\cal B} - 4 m_ba K_{\sigma\cdot F} C_{\sigma\cdot F},
	\label{eq:zBlat=zB}
\end{eqnarray}
where the short-distance coefficient $C_{\sigma\cdot F}$ appears in the
relation
\begin{equation}
	\bar{b}i\sigma^{\mu\nu}F_{\mu\nu}b \doteq
		-2 C_{\sigma\cdot F} {\cal O}_{\cal B}.
\end{equation}
At the tree level, $C_{\sigma\cdot F}^{[0]}=1$.
For the [axial] vector current,
one inserts Eq.~(\ref{eq:VcontHQET}) [Eq.~(\ref{eq:AcontHQET})]
into Eq.~(\ref{eq:LEV}) [Eq.~(\ref{eq:LEA})],
neglects terms of order~$m^2a^2$,
and compares with Eq.~(\ref{eq:VlatHQET}) [Eq.~(\ref{eq:VlatHQET})].
One also must match the tensor and pseudoscalar bilinears to HQET
at the dimension-three level,
\begin{eqnarray}
	\bar{q}i\sigma^{\mu\nu}b & \doteq &
		C_{T_+} \eta^\mu_\alpha \eta^\nu_\beta
			\bar{q}i\sigma^{\alpha\beta}h_v - 
		C_{T_-} \bar{q}(v^\mu i\gamma_\perp^\nu -
			v^\nu i\gamma_\perp^\mu)h_v, \\
	\bar{q}i\gamma_5b & \doteq & C_P \bar{q}i\gamma_5h_v,
\end{eqnarray}
with short-distance coefficients $C_{T_\pm}$ and~$C_P$.
At the tree level, $C_{T_\pm}^{[0]}=C_P^{[0]}=1$.
After carrying out these steps, one finds that 
\begin{eqnarray}
	Z_{V_\parallel}                     & = & Z_V , \label{eq:ZVpara} \\
	Z_{V_\perp}^{-1}                    & = & Z_V^{-1} +
		(m_q+m_b)a K_VC_{T_-}/C_{V_\perp} , \label{eq:ZVperp} \\
	Z_{V_\parallel} B_{V1}^{\text{lat}} & = & B_{V1} + a Z_V K_VC_{T_-} ,
		\label{eq:ZVB1} \\
	Z_{V_\parallel} B_{Vi}^{\text{lat}} & = & B_{Vi} + a Z_V K_VC_{T_+} ,
		\quad i=2,6 , \label{eq:ZVB26} \\
	Z_{V_\parallel} B_{V3}^{\text{lat}} & = & B_{V3} - a Z_V K_VC_{T_+} ,
		\label{eq:ZVB3}  \\
	Z_{V_\parallel} B_{V4}^{\text{lat}} & = & B_{V4} - a Z_V K_VC_{T_-} ,
		\label{eq:ZVB4} \\
	Z_{V_\parallel} B_{V5}^{\text{lat}} & = & B_{V5} -
		a Z_V K_V(C_{T_+}-C_{T_-}) \label{eq:ZVB5}
\end{eqnarray}
from matching the vector current, and
\begin{eqnarray}
	Z_{A_\perp}                         & = & Z_A , \label{eq:ZAperp} \\
	Z_{A_\parallel}^{-1}                & = & Z_A^{-1} +
		(m_q+m_b)a K_AC_P/C_{A_\parallel} , \label{eq:ZApara} \\
	Z_{A_\perp}     B_{Ai}^{\text{lat}} & = & B_{Ai} + O(a^2),    
		\quad i=1,2,5 , \label{eq:ZAB125} \\
	Z_{A_\perp}     B_{Ai}^{\text{lat}} & = & B_{Ai} + a Z_A K_AC_P ,
		\quad i=3,6 ,   \label{eq:ZAB36}  \\
	Z_{A_\perp}     B_{A4}^{\text{lat}} & = & B_{A4} - a Z_A K_AC_P
		\label{eq:ZAB4}
\end{eqnarray}
from matching the axial vector current.
Of course, these relations hold only when describing the same lattice
currents~$V^\mu_{\text{lat}}$ and~$A^\mu_{\text{lat}}$, and then only
to order~$a^2$.
Considering similar relations for the whole tower of higher-dimension
operators, one sees
\begin{eqnarray}
	\lim_{a\to 0} {\cal C}_{\cal O}^{\text{lat}} & = &
		{\cal C}_{\cal O}, \\
	\lim_{a\to 0} Z_{Ji} B_{Ji}^{\text{lat}} & = & B_{Ji}.
\end{eqnarray}
Eqs.~(\ref{eq:ZVperp})--(\ref{eq:ZVB5})
 and~(\ref{eq:ZApara})--(\ref{eq:ZAB4}) illustrate for the
next-to-leading dimension operators how the limit is accelerated for
standard $O(a)$ improvement, with $K_{\sigma\cdot F}$, $K_V$, and $K_A$
themselves of order~$a$.

Equations~(\ref{eq:ZVpara})--(\ref{eq:ZAB4}) show that HQET matching
connects smoothly to Symanzik matching in the limit where both apply.
HQET matching is, therefore, a natural and attractive extension
into the more practical region where $ma$ is not very small.
Continuum QCD still can be approximated well, but now order by order
in the heavy-quark expansion.

The remainder of this paper pursues this program in perturbation
theory.
One-loop corrections to the rest mass~$m_1$ and the kinetic mass~$m_2$
have been considered already in Ref.~\cite{Mertens:1998wx}.
The one-loop correction to ${\cal C}_{\cal B}$ would require a
generalization of the calculation of $K_{\sigma\cdot F}$~\cite{Wohlert:1987rf}
to incorporate the full mass dependence of the quark-gluon vertex.
In this paper we focus on heavy-light currents.
We construct lattice currents suitable for matching through order
$1/m_Q$ in the heavy quark expansion.
We then calculate the matching factors $Z_{J_\parallel}$ and
$Z_{J_\perp}$ at the one-loop level, which are needed to fix the overall
normalization of the heavy-light currents.
Currents suitable for heavy-to-heavy transitions $b\to c$ are considered
in a companion paper~\cite{Kronfeld:1999tk}.

\section{Lattice Action and Currents}
\label{sec:lattice}

In this section our aim is to define heavy-light currents with Wilson
fermions that are suited to the HQET matching formalism.
Because Wilson fermions have the right particle content and obey the
heavy-quark symmetries, the descriptive part of the formalism applies
in any case.
To use HQET to match lattice gauge theory to continuum QCD, however,
we would like to ensure that $\delta{\cal C}_i$ and $\delta B_{Ji}$
[cf.\ Eqs.~(\ref{eq:deltaC}) and (\ref{eq:deltaB})] remain bounded in
the infinite-mass limit.
Good behavior is attained by mimicking the structure of
Eqs.~(\ref{eq:Q1})--(\ref{eq:Q6}), so that improvement terms are
guaranteed to remain small.
Then we would like to adjust free parameters in the currents so that
$\delta{\cal C}_i$ and $\delta B_{Ji}$ (approximately) vanish.
We show how to do so in perturbation theory, obtaining
$B^{\text{lat}}_{Ji}$ at the tree level and, in Sec.~\ref{sec:loop},
the matching factors $Z_{J_\parallel}$ and $Z_{J_\perp}$ at the
one-loop level.

A suitable lattice Lagrangian was introduced in
Ref.~\cite{El-Khadra:1997mp}.
It is convenient to write the lattice Lagrangian
${\cal L}_{\text{lat}}={\cal L}_0+{\cal L}_B+{\cal L}_E$.
The first term is
\begin{eqnarray}
	{\cal L}_0	&=& - (m_0+m_{0\text{cr}})
			\bar{\psi}(x)\psi(x)
		- 
			\case{1}{2} \bar{\psi}(x)\left[
			(1+\gamma_4){D_4^-}_{\text{lat}} -
			(1-\gamma_4){D_4^+}_{\text{lat}}\right]\psi(x)
		\label{eq:S0} \\ &-&
			\zeta \bar{\psi}(x)
			\bbox{\gamma}\cdot\bbox{D}_{\text{lat}} \psi(x) 
		 +	\case{1}{2} r_s\zeta a
			\bar{\psi}(x)\triangle^{(3)}_{\text{lat}}\psi(x).
	\nonumber
\end{eqnarray}
The mass counterterm $m_{0\text{cr}}$ is included here so that,
by definition, $m_0=0$ for massless quarks.
The covariant difference operators ${D_4^\pm}_{\text{lat}}$,
$\bbox{D}_{\text{lat}}$, and $\triangle^{(3)}_{\text{lat}}$,
are defined in Ref.~\cite{El-Khadra:1997mp}.
They carry the label ``lat'' to distinguish them from the
continuum covariant derivatives in Secs.~\ref{subsec:sym}
and~\ref{subsec:hqetcont}.
The symbol $\psi$ is reserved in this paper for lattice fermion fields.
The temporal kinetic term is conventionally normalized, but the
spatial kinetic term is multiplied with the coupling~$\zeta$.
The coupling $r_s$ is, in the technical sense,
redundant~\cite{El-Khadra:1997mp}, but is included to solve the
doubling problem\cite{Wilson:1975hf}.

For ${\cal L}_0$ the tree-level relations between its couplings and
the coefficients in the~${\cal L}_{\text{HQET}}$ are well known.
By matching the kinetic energy, one finds (for $\bbox{v}=\bbox{0}$)
\begin{equation}
	{{\cal C}^{\text{lat}}_2}^{[0]}        =
		\frac{1}{2m^{[0]}_2a} =
		\frac{\zeta^2}{m_0a(2+m_0a)} +
		\frac{r_s\zeta}{2(1+m_0a)}. \label{eq:m2}
\end{equation}
At higher orders in perturbation theory, ${\cal C}_2$ remains
(for $\bbox{v}=\bbox{0}$) the kinetic mass of the quark, which is
expressed in terms of the self energy in Ref.~\cite{Mertens:1998wx}.

${\cal L}_0$ has cutoff artifacts, which are described
by dimension-five and -higher operators
in ${\cal L}_{\text{Sym}}$  (if $m_qa\ll1$)
or~${\cal L}_{\text{HQET}}$ (if $m_Q\gg\Lambda_{\text{QCD}}$).
The dimension-five effect can be reduced by adding
\begin{eqnarray}
	{\cal L}_B & = & \case{i}{2} a c_B\zeta \,
		\bar{\psi}(x)\bbox{\Sigma}\cdot\bbox{B}_{\text{lat}}(x)\psi(x),
	\label{eq:SB} \\
	{\cal L}_E & = & \case{1}{2} a c_E\zeta \,
		\bar{\psi}(x)\bbox{\alpha}\cdot\bbox{E}_{\text{lat}}(x)\psi(x),
	\label{eq:SE}
\end{eqnarray}
and suitably adjusting of~$c_B$ and~$c_E$.
The lattice chromomagnetic and chromoelectric fields,
$\bbox{B}_{\text{lat}}$ and $\bbox{E}_{\text{lat}}$,
are those given in Ref.~\cite{El-Khadra:1997mp}.

By matching the gluon-quark vertex, one finds
\begin{equation}
	{{\cal C}^{\text{lat}}_{\cal B}}^{[0]} =
		\frac{1}{2m^{[0]}_{\cal B}a} =
		\frac{\zeta^2}{m_0a(2+m_0a)} +
		\frac{c_B\zeta}{2(1+m_0a)}. \label{eq:mB}
\end{equation}
Higher-order corrections to ${\cal C}^{\text{lat}}_{\cal B}$
have not been obtained.
By comparing Eqs.~(\ref{eq:m2}) and~(\ref{eq:mB}) one sees, however,
that $c_B=r_s+O(g^2)$ is needed to adjust
${\cal C}^{\text{lat}}_{\cal B}$ to its continuum
counterpart~${\cal C}_{\cal B}=z_{\cal B}/2m_2$.

The Euclidean action is $S=-a^4\sum_x{\cal L}(x)$.
Special cases are the Wilson action~\cite{Wilson:1975hf}, which
sets $r_s=\zeta=1$, $c_B=c_E=0$; and the Shei\-kholeslami-Wohlert
action~\cite{Sheikholeslami:1985ij}, which sets $r_s=\zeta=1$,
$c_B=c_E\equiv c_{\text{SW}}$.
But to remove lattice artifacts for arbitrary masses,
the couplings~$r_s$, $\zeta$, $c_B$ and~$c_E$ must be taken to depend
on $m_0a$~\cite{El-Khadra:1997mp}.
Our analytical results for the integrands of Feynman diagrams, given in
Appendix~\ref{app:dirac}, are for arbitrary choices of these couplings.
Indeed, our expressions allow the heavy and light quarks to have
different values of all couplings.

Heavy-light currents are defined in an essentially similar way.
For convenience, first define a ``rotated''
field~\cite{Kronfeld:1995nu,El-Khadra:1997mp}
\begin{equation}
	\Psi_q = \left[1 + ad_1
		\bbox{\gamma}\cdot\bbox{D}_{\text{lat}}
		\right] \psi_q,
	\label{eq:rotate}
\end{equation}
where $\psi_q$ is the field in ${\cal L}_0$ of flavor~$q$, and
$\bbox{D}_{\text{lat}}$ is again the symmetric covariant difference
operator.
Simple bilinears with the right quantum numbers are
\begin{eqnarray}
	V^\mu_0 & = & \bar{\Psi}_qi\gamma^\mu \Psi_b,
	\label{eq:Vlat} \\
	A^\mu_0 & = & \bar{\Psi}_qi\gamma^\mu\gamma_5 \Psi_b.
	\label{eq:Alat}
\end{eqnarray}
The subscript ``0'' implies that, as with ${\cal L}_0$, some improvement
is desired.
To ensure a good large-$ma$ limit, one should pattern the improved
current after the right-hand side of Eq.~(\ref{eq:VlatHQET}).
Thus, we take
\begin{eqnarray}
	V^\mu_{\text{lat}} & = & V^\mu_0 -
		\sum_{i=1}^6 b_{Vi} Q^\mu_{Vi},
	\label{eq:Vimp} \\
	A^\mu_{\text{lat}} & = & A^\mu_0 -
		\sum_{i=1}^6 b_{Ai} Q^\mu_{Ai},
	\label{eq:Aimp}
\end{eqnarray}
where the $b_{Ji}$ are adjustable,
and the dimension-four lattice operators are
\begin{eqnarray}
	Q^\mu_{V1} & = & - v^\mu \bar{\psi}_q
		i{\kern+0.1em /\kern-0.55em v}
		{\kern+0.1em /\kern-0.65em D_\perp}_{\text{lat}} \psi_b,
		\label{eq:Q1lat} \\
	Q^\mu_{V2} & = & \bar{\psi}_qi\gamma^\mu_\perp
		{\kern+0.1em /\kern-0.65em D_\perp}_{\text{lat}} \psi_b,
		\label{eq:Q2lat} \\
	Q^\mu_{V3} & = &
		\bar{\psi}_q i {D^\mu_\perp}_{\text{lat}} \psi_b,
		\label{eq:Q3lat} \\
	Q^\mu_{V4} & = & - v^\mu \bar{\psi}_q
		{\kern+0.1em /\kern-0.65em \loarrow{D}_\perp}_{\text{lat}}
		i{\kern+0.1em /\kern-0.55em v} \psi_b, \label{eq:Q4lat} \\
	Q^\mu_{V5} & = & \bar{\psi}_q
		{\kern+0.1em /\kern-0.65em \loarrow{D}_\perp}_{\text{lat}}
		i\gamma^\mu_\perp \psi_b,  \label{eq:Q5lat} \\
	Q^\mu_{V6} & = &
		\bar{\psi}_qi {\loarrow{D}^\mu_\perp}_{\text{lat}} \psi_b,
		\label{eq:Q6lat}
\end{eqnarray}
and each lattice operator $Q^\mu_{Ai}$ is obtained from $Q^\mu_{Vi}$
by replacing $\bar{\psi}_q$ with~$-\bar{\psi}_q\gamma_5$.
Lattice quark fields do not satisfy Eq.~(\ref{eq:Pvh}), so
${\kern+0.1em /\kern-0.55em v}$ appears explicitly.
In practice, one uses the rest frame here, $v=(i,\bbox{0})$, as in
Eq.~(\ref{eq:rotate}).
An analogous construction for lattice NRQCD has been given by
Morningstar and Shigemitsu~\cite{Morningstar:1998ep}.

It is worthwhile to emphasize the difference between
Eqs.~(\ref{eq:VlatHQET}) and~(\ref{eq:Vimp}).
Equation~(\ref{eq:VlatHQET}) is a general HQET description of any
heavy-light lattice current.
Equation~(\ref{eq:Vimp})  is a definition of a specific lattice
current, namely the one used in this paper (and in calculations of
$f_B$ and other hadronic matrix elements).
In the same vein, the ${\cal Q}_{Ji}$ in Eqs.~(\ref{eq:Q1})--(\ref{eq:Q6})
are HQET operators, whereas the $Q_{Ji}$ in
Eqs.~(\ref{eq:Q1lat})--(\ref{eq:Q6lat}) are lattice operators.
Finally, the coefficients $B^{\text{lat}}_{Ji}$ are the output of
a matching calculation: they depend on the~$b_{Ji}$, which must be
adjusted to make~$\delta B_{Ji}$ vanish.

To illustrate, let us consider the calculation of the
coefficients~$B^{\text{lat}}_{Ji}$ at the tree level.
One computes on-shell matrix elements such as
$\langle q|J_{\text{lat}}|b\rangle$ and
$\langle 0|J_{\text{lat}}|\bar{q}b\rangle$ in lattice gauge theory and
compares them to the corresponding matrix elements in HQET.
Then one finds
\begin{eqnarray}
	{C^{\text{lat}}_{J_\parallel}}^{[0]} & = &
		{C^{\text{lat}}_{J_\perp}}^{[0]} =
		e^{-(m^{[0]}_{1q}+m^{[0]}_{1b})a/2} ,
	\label{eq:matchC[0]} \\
	{B^{\text{lat}}_{Ji}}^{[0]} & = &
		e^{-(m^{[0]}_{1q}+m^{[0]}_{1b})a/2} \left(
		\frac{1}{2m^{[0]}_3} + b^{[0]}_{Ji}\right), \quad i\le2
	\label{eq:B1B2lat} \\
	{B^{\text{lat}}_{Ji}}^{[0]} & = &
		e^{-(m^{[0]}_{1q}+m^{[0]}_{1b})a/2} b^{[0]}_{Ji},
		\quad i\ge 3 \label{eq:B3-6lat}
\end{eqnarray}
where
\begin{equation}
	m_1^{[0]}a = \ln(1 + m_0a)
\end{equation}
and, for our lattice Lagrangian and currents,
\begin{equation}
	\frac{1}{2m^{[0]}_3a} = \frac{\zeta(1+m_0a)}{m_0a(2+m_0a)} - d_1.
 	\label{eq:m3}
\end{equation}
Since (continuum QCD's) $C_J^{[0]}=1$ there already is a non-trivial
matching factor at the tree level relating the lattice and continuum
currents,
$Z_{J_\parallel}^{[0]}=Z_{J_\perp}^{[0]}=e^{(m_{1q}+m_{1b})a/2}$.

After comparing Eqs.~(\ref{eq:B1B2lat})--(\ref{eq:B3-6lat}) with
Eqs.~(\ref{eq:B1B2})--(\ref{eq:B3-6}), one sees that
one can take $b_{Ji}^{[0]}=0$ for all six operators,
if $d_1$ is adjusted correctly.
At the tree level, the way to adjust $d_1$ is to set $m^{[0]}_3$ equal
to the (tree-level) heavy-quark mass.
In the effective Lagrangian there are two quark masses, the rest
mass~$m_1$ and the kinetic mass~$m_2$.
The former has no effect on matrix elements (and a trivial, additive
effect on the mass spectrum).
As discussed above, heavy-quark cutoff effects in matrix elements are
reduced if ${\cal C}^{\text{lat}}_2={\cal C}_2$, which means one should
identify the continuum quark mass with the kinetic mass.
Thus, one should set $m^{[0]}_3=m^{[0]}_2$, which is obtained if
one adjusts
\begin{equation}
	d_1 = \frac{\zeta(1+m_0a-\zeta)}{m_0a(2+m_0a)} - 
		\frac{r_s\zeta}{2(1+m_0a)}.
	\label{eq:d1(m)}
\end{equation}
The same rotation also improves heavy-heavy currents at the tree level.

Beyond the tree level, it is convenient to define $d_1$ for
the spatial component of the degenerate-mass, heavy-heavy
current~\cite{Kronfeld:1999tk}.
Then the corrections heavy-heavy current analogous to $Q_{V2}$
and~$Q_{V5}$ would be superfluous, but for unequal masses they are
still required.

For equal mass currents it is possible to compute $Z_{V_\parallel}$
nonperturbatively for all masses~$m_b$.
One may therefore prefer to
write~\cite{Hashimoto:2000yp,Simone:2000nv,El-Khadra:2001rv}
\begin{equation}
	Z_{J_{\parallel,\perp}^{ub}} =
		\sqrt{ Z_{V_\parallel^{uu}} Z_{V_\parallel^{bb}} }
		\rho_{J_{\parallel,\perp}^{ub}}
	\label{eq:rho}
\end{equation}
and compute only the factor $\rho_{J_{\parallel,\perp}^{ub}}$ in
perturbation theory.
To calculate the pre-factor $Z_{V_\parallel^{bb}}$ appearing
in Eq.~(\ref{eq:rho}), one must have a massive quark in the final
state.
The definition of the heavy-heavy matching factor is given in our
companion paper~\cite{Kronfeld:1999tk},
along with a calculation of its one-loop level contribution.
We give the results for heavy-light~$\rho_{J_{\parallel,\perp}}$
in Sec.~\ref{sec:loop}.

For a light quark, with $m_qa\ll 1$, the right hand side of
Eq.~(\ref{eq:d1(m)}) vanishes linearly in~$m_0a$.
Therefore, $1-m_{3q}/m_{2q}$ is $O(m_q^2a^2)$, and the distinction
between $m_{3q}$ and $m_{2q}$ is negligible.
For this reason, and to simplify calculation, we set $m_q=0$.
Then Eq.~(\ref{eq:d1(m)}) implies $d_1=0$ for the light quark.

\section{One-Loop Results}
\label{sec:loop}

In this section we present results for the matching factors at the
one-loop level in perturbation theory.
The one-loop contributions are known for the
Wilson~\cite{Kuramashi:1998tt} and Sheikholeslami-Wohlert (SW)
actions~\cite{Ishikawa:1997xh}.
Both these works omit the rotation term in the
current~\cite{Kronfeld:1995nu,El-Khadra:1997mp},
which is needed to obtain $1/m_3$ correctly.
In this section we complete the work started in
Ref.~\cite{Ishikawa:1997xh} and report results with the clover term
and with the rotation.
For comparison we also present our results without the rotation,
both with and without the clover term.

The computer code for generating these results is freely
available~\cite{p:epaps}.

The matching factors $Z_J$ ($J=V_\parallel$, $V_\perp$, $A_\parallel$,
and $A_\perp$) are simply the ratios of the lattice and continuum
radiative corrections:
\begin{equation}
	Z_J = \frac{%
	\left[Z_{2h}^{1/2}\Lambda_J Z_{2l}^{1/2}\right]^{\rm cont}}{%
	\left[Z_{2h}^{1/2}\Lambda_J Z_{2l}^{1/2}\right]^{\rm lat~}},
	\label{eq:ZJGamma}
\end{equation}
where $Z_{2h}$ and $Z_{2l}$ are wave-function renormalization factors of
the heavy and light quarks, and the vertex function $\Lambda_J$ is
the sum of one-particle irreducible three-point diagrams, in which
one point comes from the current~$J$ and the other two from the
external quark states.

The expression relating $Z_2$ to the lattice self energy, for all masses
and gauge couplings, can be found in Ref.~\cite{Mertens:1998wx}.
Its dominant mass dependence is
\begin{equation}
	Z_2 \propto e^{-m_1a},
\end{equation}
where $m_1$ is the all-orders rest mass (of the heavy quark).
This mass dependence is not present in the vertex function or the 
continuum part of Eq.~(\ref{eq:ZJGamma}).
Consequently, we write
\begin{equation}
	e^{-m^{[0]}_1a/2} Z_{J_\Gamma} =
		1 + \sum_{l=1}^\infty g_0^{2l} Z_{J_\Gamma}^{[l]},
	\label{eq:ZJ[1]}
\end{equation}
so that the $Z_{J_\Gamma}^{[l]}$ are only mildly mass dependent.
(A~slightly different convention was used in presenting results for 
$Z_2$ in Ref.~\cite{Mertens:1998wx}.)
By construction, this mass dependence in $\rho_{J_\Gamma}$ cancels out
in a gauge-invariant, all orders way.
So, we write
\begin{equation}
	\rho_{J_\Gamma} =
		1 + \sum_{l=1}^\infty g_0^{2l} \rho_{J_\Gamma}^{[l]}.
	\label{eq:rhoJ[1]}
\end{equation}

This rest of this section is split into two subsections.
In the first, we present our results for the full mass dependence of
$Z_{J_\Gamma}^{[1]}$ and~$\rho_{J_\Gamma}^{[1]}$.
In the second, we discuss the related calculation of the
Brodsky-Lepage-Mackenzie scale~$q^*$.
In both cases, we discuss fully a range of checks on our calculations.

\subsection{$Z_J^{[1]}$ and $\rho_J^{[1]}$}

The combinations of wave-function and vertex renormalization
in~$Z_J$ are gauge invariant and ultraviolet and infrared
finite.
For vanishing light quark mass there is a collinear divergence (which
can be regulated by an infinitesimally small mass), but it is common
to lattice and continuum functions.
In the desired ratio~(\ref{eq:ZJGamma}), the divergence cancels, and
the result is independent of the scheme for regulating the collinear
singularity.
For large $ma$ a remnant of this cancellation appears.
The lattice theory approaches its static limit, where its ultraviolet
behavior is non-logarithmic.
But the region of momentum $a^{-1}<q<m$ in the continuum diagrams
generates logarithms.
At the one-loop level one must find $3\ln(ma)$, with the same
anomalous dimension as in Eqs.~(\ref{eq:Cparallel})
and~(\ref{eq:Cperp}).
At higher loops the usual polynomial in $\ln(ma)$ will arise.

We have calculated the one-loop Feynman diagrams for the action
specified in Eqs.~(\ref{eq:S0})--(\ref{eq:SE}), with arbitrary
$m_0$,    $r_s$,  $\zeta$,  $c_B$,  and $c_E$
for the incoming heavy quark,
$m'_0=0$, $r'_s$, $\zeta'$, $c'_B$, and $c'_E$
for the outgoing light quark.
The needed Feynman rules are in Ref.~\cite{Mertens:1998wx}, apart from
three new rules for the current itself,
which are in Appendix~\ref{app:feynman}.
As shown in Appendix~\ref{app:dirac}, we have found a simple way to
incorporate the rotation into the Dirac algebra.
The resulting analytical expressions are surprisingly compact,
and they are given explicitly in Appendices~\ref{app:dirac}
and~\ref{app:useful}.

\begin{figure}
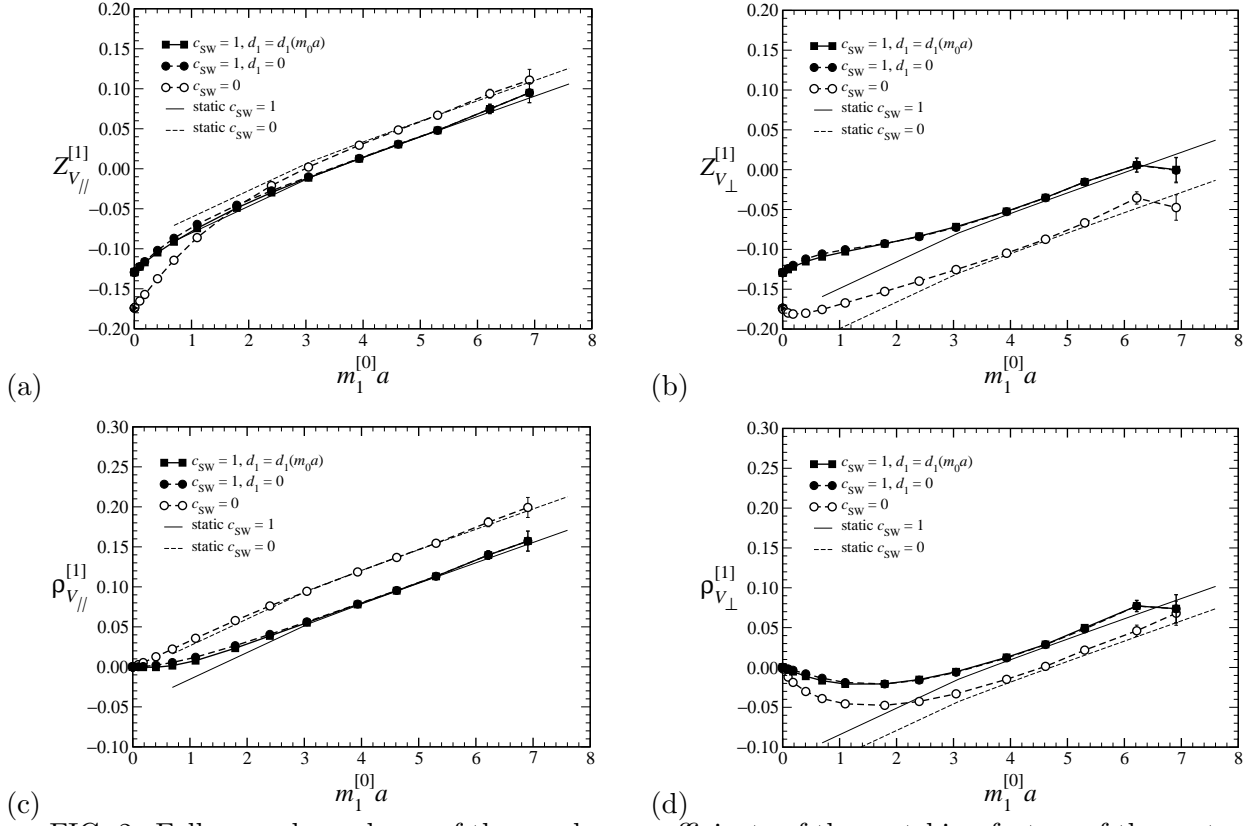

	\centering\small
	(a) \includegraphics[width=0.44\textwidth]{ZV4m1.eps}   \hfill
	(b) \includegraphics[width=0.44\textwidth]{ZVjm1.eps}   \\[0.5em]
	(c) \includegraphics[width=0.44\textwidth]{rhoV4m1.eps} \hfill
	(d) \includegraphics[width=0.44\textwidth]{rhoVjm1.eps}
	\caption[fig:ZV]{Full mass dependence of the one-loop coefficients
	of the matching factors of the vector current
	   (a)~$Z^{[1]}_{V_\parallel}$,        (b)~$Z^{[1]}_{V_\perp}$,
	(c)~$\rho^{[1]}_{V_\parallel}$, and (d)~$\rho^{[1]}_{V_\perp}$.
	Filled (open) symbols denote the SW (Wilson) action;
	solid (dotted) lines connecting squares (circles) 
	indicate the rotation is included (omitted).}
	\label{fig:ZV}
\end{figure}

\begin{figure}
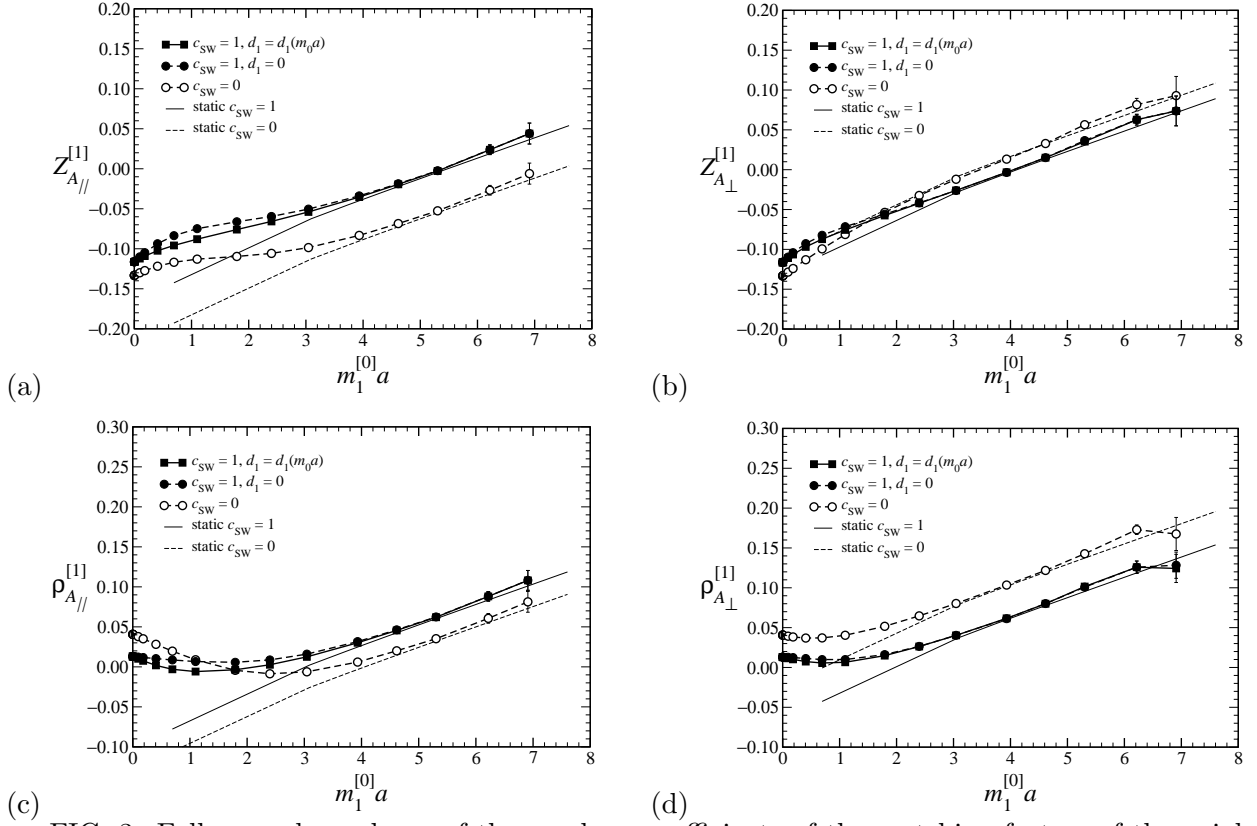

	\centering\small
	(a) \includegraphics[width=0.44\textwidth]{ZA4m1.eps} \hfill
	(b) \includegraphics[width=0.44\textwidth]{ZAjm1.eps} \\[0.5em]
	(c) \includegraphics[width=0.44\textwidth]{rhoA4m1.eps} \hfill
	(d) \includegraphics[width=0.44\textwidth]{rhoAjm1.eps}
	\caption[fig:ZA]{Full mass dependence of the one-loop coefficients
	of the matching factors of the axial vector current
	   (a)~$Z^{[1]}_{A_\parallel}$,        (b)~$Z^{[1]}_{A_\perp}$,
	(c)~$\rho^{[1]}_{A_\parallel}$, and (d)~$\rho^{[1]}_{A_\perp}$.}
	\label{fig:ZA}
\end{figure}

We have evaluated these expressions for $r_s=\zeta=1$ and
$c_E=c_B\equiv c_{\text{SW}}$.
Thus, the numerical results correspond to the SW action
($c_{\text{SW}}=1$) and to the Wilson action ($c_{\text{SW}}=0$).
Figure~\ref{fig:ZV} plots the full mass dependence of
the matching factors for the vector current,
(a)~$Z_{V_\parallel}$,     (b)~$Z_{V_\perp}$,
(c)~$\rho_{V_\parallel}$, and (d)~$\rho_{V_\perp}$.
These numerical results are for the SW action with rotation (solid
lines) and also for the SW and Wilson actions without the rotation
(dotted lines).
Figure~\ref{fig:ZA} plots the full mass dependence of
the matching factors for the axial vector current,
(a)~$Z_{A_\parallel}$,     (b)~$Z_{A_\perp}$,
(c)~$\rho_{A_\parallel}$, and (d)~$\rho_{A_\perp}$.
These and the following figures are plotted against $m_1^{[0]}a$ because
this variable conveniently covers the whole mass range: for small mass
$m_1\approx m_2$, and for large mass $m_1a\approx\ln m_2a$.

We have carried out several checks on our calculations.
In each case, identical numerical results have been obtained with two
or more completely independent programs.
The results for $Z_{J_{\parallel,\perp}}$ agree with those previously
obtained, for $c_{\text{SW}}=0$~\cite{Kuramashi:1998tt} and for
$c_{\text{SW}}=1$, $d_1=0$~\cite{Ishikawa:1997xh}.
We have also reproduced limiting cases, as we briefly discuss below.

For $ma=0$ our calculation reduces to the usual matching calculation
for massless quarks.
We find (with $C_F=4/3$)
\begin{eqnarray}
	Z_{V_\parallel}^{[1]} = Z_{V_\perp}^{[1]} & = & \left\{
		\begin{array}{ll}
			- 0.129423(6), & c_{\text{SW}} = 1, \\
			- 0.174073(7), & c_{\text{SW}} = 0,
		\end{array} \right. \\
	Z_{A_\parallel}^{[1]} = Z_{A_\perp}^{[1]} & = & \left\{
		\begin{array}{ll}
			- 0.116450(5), & c_{\text{SW}} = 1, \\
			- 0.133365(5), & c_{\text{SW}} = 0,
		\end{array} \right.
\end{eqnarray}
in excellent agreement with previous work for
$c_{\text{SW}}=1$ \cite{Gabrielli:1991us,Luscher:1996ax,Capitani:2001xi} and
$c_{\text{SW}}=0$~\cite{Capitani:2001xi,Martinelli:1983mw,Bernard:1999sx}.
(Reference~\cite{Capitani:2001xi} gives precise results as
a polynomial in~$c_{\text{SW}}$.)

As the mass tends to infinity, these actions and currents all
lead, up to an unphysical factor, to the same vertices and quark
propagator---a Wilson line.
Perturbative corrections to the vertex functions must respect this
universal static limit, and, therefore, they must tend to a universal 
value.
As $ma\to\infty$, one expects the $Z$ factors for a massive quark to
approach those for the static limit, namely
\begin{equation}
	Z_J^{[1]} = \frac{C_F}{16\pi^2}
		\left[\gamma_h\ln(m_2a)^2 + z_J^{[1]}\right],
\end{equation}
where the constant $z_J^{[1]}$ depends on the current~$J$ and
on~$c_{\text{SW}}$ (of the light quark).
Since $\ln(m_2a)\approx m_1a$ in this region one expects the linear
behavior seen in Figs.~\ref{fig:ZV} and~\ref{fig:ZA}.
The static limit is also shown in Figs.~\ref{fig:ZV} and~\ref{fig:ZA}
with
\begin{eqnarray}
	z_{V_\parallel}^{[1]} & = & \left\{
		\begin{array}{rl}
			- 10.248, & c_{\text{SW}} = 1, \\
			-  7.929, & c_{\text{SW}} = 0,
		\end{array} \right. \\
	z_{A_\perp}^{[1]}     & = & \left\{
		\begin{array}{rl}
			- 12.248, & c_{\text{SW}} = 1, \\
			-  9.929, & c_{\text{SW}} = 0,
		\end{array} \right. \\
	z_{V_\perp}^{[1]}     & = & \left\{
		\begin{array}{rl}
			- 18.414, & c_{\text{SW}} = 1, \\
			- 24.379, & c_{\text{SW}} = 0,
		\end{array} \right. \\
	z_{A_\parallel}^{[1]} & = & \left\{
		\begin{array}{rl}
			- 16.414, & c_{\text{SW}} = 1, \\
			- 22.379, & c_{\text{SW}} = 0.
		\end{array} \right. 
\end{eqnarray}
We have obtained these constants ourselves~\cite{Ishikawa:1998rv}.
They agree with previous (less precise) results for 
$c_{\text{SW}}=1$~\cite{Borrelli:1992fy} and 
$c_{\text{SW}}=0$~\cite{Eichten:1990zv}.
As one can see from looking at Figs.~\ref{fig:ZV} and~\ref{fig:ZA}, the
static result is a good approximation for $m_1^{[0]}a>5$
or, equivalently, $m_0a\approx m_2a>150$.

Some of the points at the highest masses have large error and lie nearly
one $\sigma$ off the curve.
The origin of this behavior is that the lattice and continuum integrals
are dominated by different momenta: the continuum integral is dominated
by the region $k\sim m_2\gg a^{-1}$, whereas the lattice integral is
dominated by the region $k\sim a^{-1}$.
This mass region is not of much practical interest, since here one
has an essentially static quark.

Equations~(\ref{eq:ZVpara})--(\ref{eq:ZAB4}) allow us to check
the small (heavy-quark) mass limit against the work of Sint and
Weisz~\cite{Sint:1997jx}.
In our conventions the matching factors~$Z_V$ and~$Z_A$ are functions
of gauge coupling and quark mass.
Thus,
\begin{eqnarray}
	Z_V(m_qa, m_ba) & = &
		Z_V \left[1 + \case{1}{2}(m_q+m_b)ab_V \right], \\
	Z_A(m_qa, m_ba) & = &
		Z_A \left[1 + \case{1}{2}(m_q+m_b)ab_A \right],
\end{eqnarray}
where, on the right-hand side, we adopt the notation of
Refs.~\cite{Luscher:1996sc,Jansen:1996ck,Sint:1997jx},
and the $Z$s and $b$s do not depend on mass.
Here only the mass dependence is displayed;
all quantities depend also on the gauge coupling.

If we omit the rotation, our currents and those considered by Sint and
Weisz coincide, apart from one-loop counterterms.
Thus, in one-loop calculations the slopes of our mass-dependent matching
factors must agree with them.
Setting $m_q=0$, and using Eqs.~(\ref{eq:ZVpara})--(\ref{eq:ZAB4}),
\begin{eqnarray}
	\frac{\partial Z^{[1]}_{V_\parallel}}{\partial m_{1b}} & = &
		\case{1}{2}b^{[1]}_V, \\
	\frac{\partial Z^{[1]}_{V_\perp}}{\partial m_{1b}} & = &
		\case{1}{2}b^{[1]}_V - K^{[1]}_V, \\
	\frac{\partial Z^{[1]}_{A_\parallel}}{\partial m_{1b}} & = &
		\case{1}{2}b^{[1]}_A - K^{[1]}_A, \\
	\frac{\partial Z^{[1]}_{A_\perp}}{\partial m_{1b}} & = &
		\case{1}{2}b^{[1]}_A.
\end{eqnarray}
To extract these slopes, we form a combination of integrands with 
three different (small) values of $m_ba$, yielding $b_J^{[1]}$ 
and~$K_J^{[1]}$ up to $O(m_ba)^2$.
In this way we find (for $c_{\text{SW}}=1$)
\begin{eqnarray}
	b_V^{[1]}    &     =     & C_F\times 0.114929(10) = 0.153239(14) \\
	             &{\rm vs.}\;& C_F\times 0.11492(4) \cite{Sint:1997jx},
		 \nonumber \\ 
	b_A^{[1]}    &     =     & C_F\times 0.114142(10) = 0.152189(14) \\
	             &{\rm vs.}\;& C_F\times 0.11414(4) \cite{Sint:1997jx},
		 \nonumber \\ 
	K_V^{[1]}    &     =     & C_F\times 0.0122499(6) = 0.016332(7) \\
	             &{\rm vs.}\;& C_F\times 0.01225(1) \cite{Sint:1997jx},
		 \nonumber \\ 
	K_A^{[1]}    &     =     & C_F\times 0.0056806(11) = 0.0075741(15) \\ 
	             &{\rm vs.}\;& C_F\times 0.005680(2) \cite{Sint:1997jx},
		 \nonumber
\end{eqnarray}
which agrees perfectly with Ref.~\cite{Sint:1997jx}.
These results have also been checked by Taniguchi and
Ukawa~\cite{Taniguchi:1998pf}.
We also obtain
\begin{equation}
	b_V^{[1]} - b_A^{[1]} = C_F\times 0.0007833(11) = 0.0010444(16)
	\label{eq:bV-bA}
\end{equation}
by subtracting the integrands first, and then integrating.
In taking the difference, large contributions from the self energy
cancel, but, even so, the near equality of~$b_V^{[1]}$ and~$b_A^{[1]}$
is a bit astonishing.
Comparing the slopes of Figs.~\ref{fig:ZV}(a) and~\ref{fig:ZA}(b) one
sees that $b_V^{[1]}-b_A^{[1]}$ for the Wilson action is not so small.

Although these checks are reassuring, the main result of this section
is to obtain the full mass dependence of the matching factors.
The results at intermediate mass, with $m_1a<3$ or, equivalently,
$m_0a<1.5$, are needed for realistic calculations of $B$ meson
properties.
This region is neither particularly close to the massless limit,
nor to the logarithmic behavior of the static limit.

\subsection{BLM scales~$q^*$}

It is well-known that perturbation theory in the bare
coupling~$g_0^2(1/a)$ converges poorly.
Therefore, we calculate the ingredients needed
to determine the Brodsky-Lepage-Mackenzie (BLM)
scale~\cite{Brodsky:1983gc,Lepage:1993xa}.
For a coupling in scheme~$S$, we denote the BLM expansion
parameter~$g_S^2(q^*_S)$.
The BLM scale~$q^*_S$ is given by
\begin{equation}
	\ln(q^*_Sa)^2 = - b_S^{(1)} +
		\frac{\int d^4k\,\ln(ka)^2\,f(ka)}{\int d^4k\,f(ka)},
	\label{eq:qstar}
\end{equation}
where $k$ is the gluon momentum, and $f(k)$ is the integrand of the
quantity of interest, e.g., $\int d^4k\,f(k)=Z_J^{[1]}$.
The constant $b_S^{(1)}$ is the $\beta_0$-dependent part of the
one-loop conversion from the arbitrary scheme~$S$ to the ``$V$~scheme'',
namely
\begin{equation}
	\frac{(4\pi)^2}{g^2_S(q)} = \frac{(4\pi)^2}{g^2_V(q)} +
		\beta_0 b_S^{(1)} + b_S^{(0)} + O(g^2),
	\label{eq:bS}
\end{equation}
where for $n_f$ light quarks $\beta_0=11-2n_f/3$, 
and $b_S^{(0)}$ is independent of~$n_f$.
The $V$-scheme coupling $g^2_V(q)$ is defined so that the Fourier
transform of the heavy-quark potential reads $V(q)=-C_Fg^2_V(q)/q^2$.
Equation~(\ref{eq:qstar}) shows that the definitions of~$q^*$ in
Refs.~\cite{Brodsky:1983gc} and~\cite{Lepage:1993xa} are identical in
the $V$~scheme.

For our matching factors it is straightforward to weight
the integrands with $\ln(ka)^2$ to obtain
\begin{equation}
	\ln(q^*_Va)^2 = \frac{{}^*Z^{[1]}}{Z^{[1]}},
	\label{eq:qStar}
\end{equation}
because the integration over $d^4k$ has no divergences.
The denominators are the one-loop coefficients given above, and the
numerators are presented now.

\begin{figure}
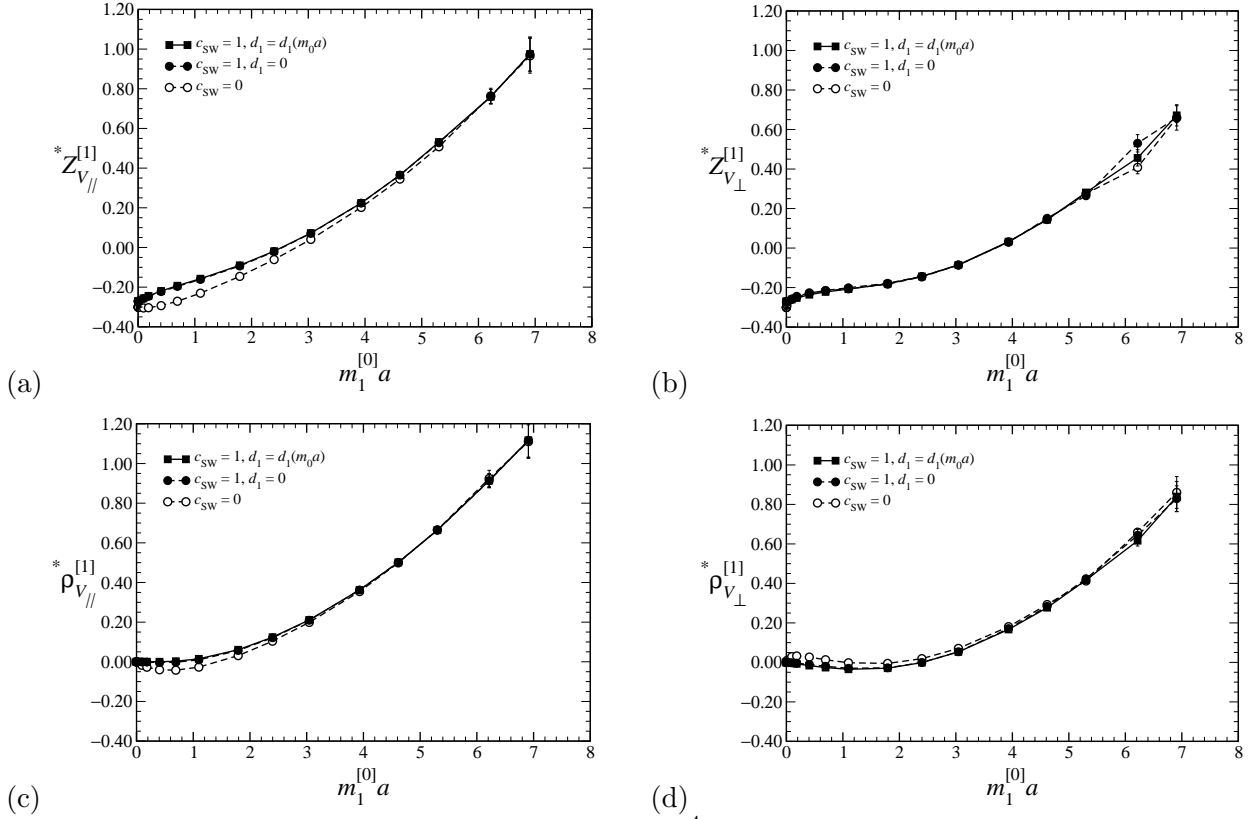

	\centering\small
	(a) \includegraphics[width=0.44\textwidth]{starZV4m1.eps} \hfill
	(b) \includegraphics[width=0.44\textwidth]{starZVjm1.eps} \\[0.5em]
	(c) \includegraphics[width=0.44\textwidth]{starrhoV4m1.eps} \hfill
	(d) \includegraphics[width=0.44\textwidth]{starrhoVjm1.eps}
	\caption[fig:starZV]{Full mass dependence of the estimated $\beta_0g^4$
	terms of the matching factors of the vector current
	   (a)~${}^*Z^{[1]}_{V_\parallel}$,        (b)~${}^*Z^{[1]}_{V_\perp}$,
	(c)~${}^*\rho^{[1]}_{V_\parallel}$, and (d)~${}^*\rho^{[1]}_{V_\perp}$.}
	\label{fig:starZV}
\end{figure}

\begin{figure}
	\centering\small
	(a) \includegraphics[width=0.44\textwidth]{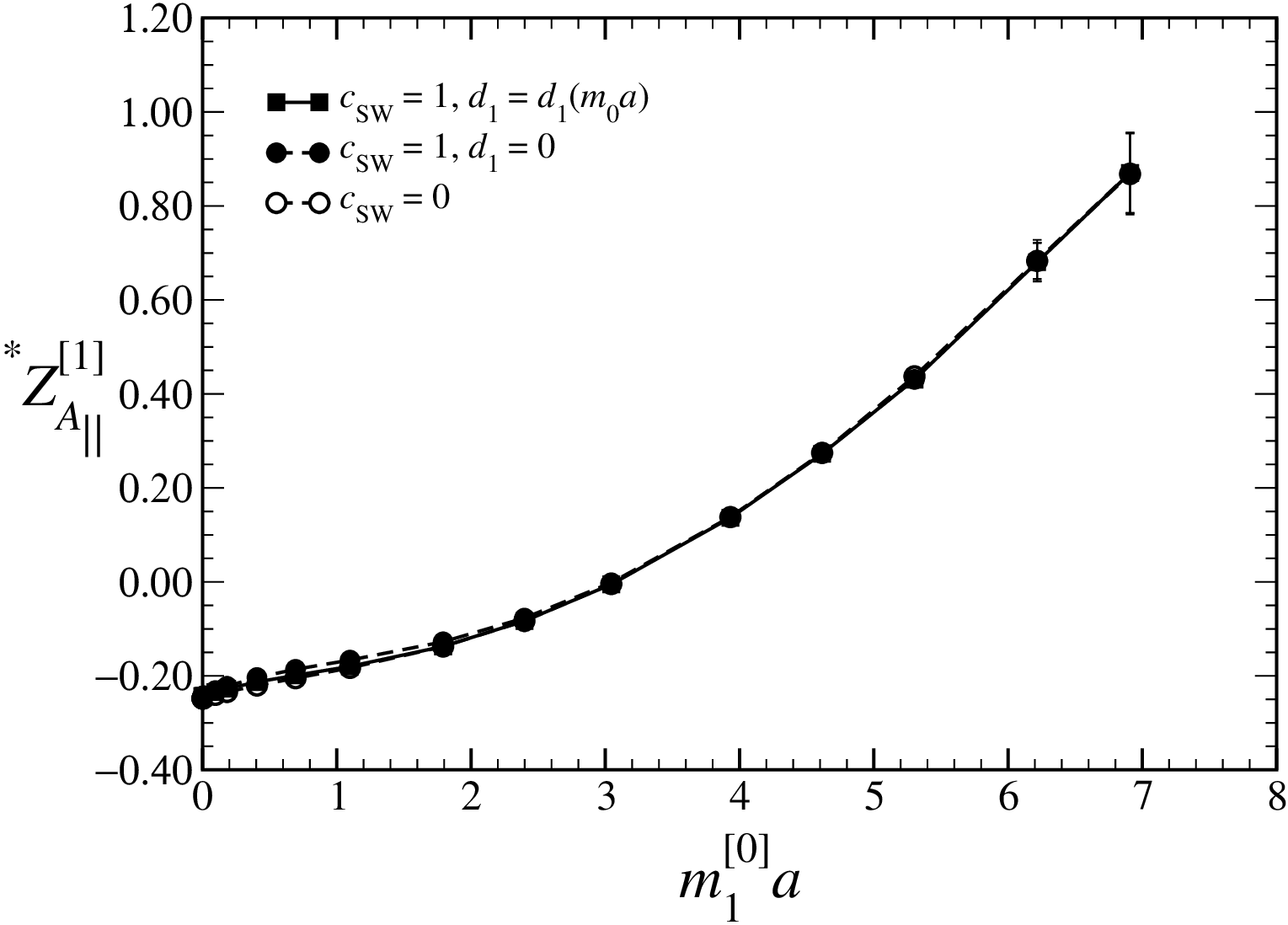} \hfill
	(b) \includegraphics[width=0.44\textwidth]{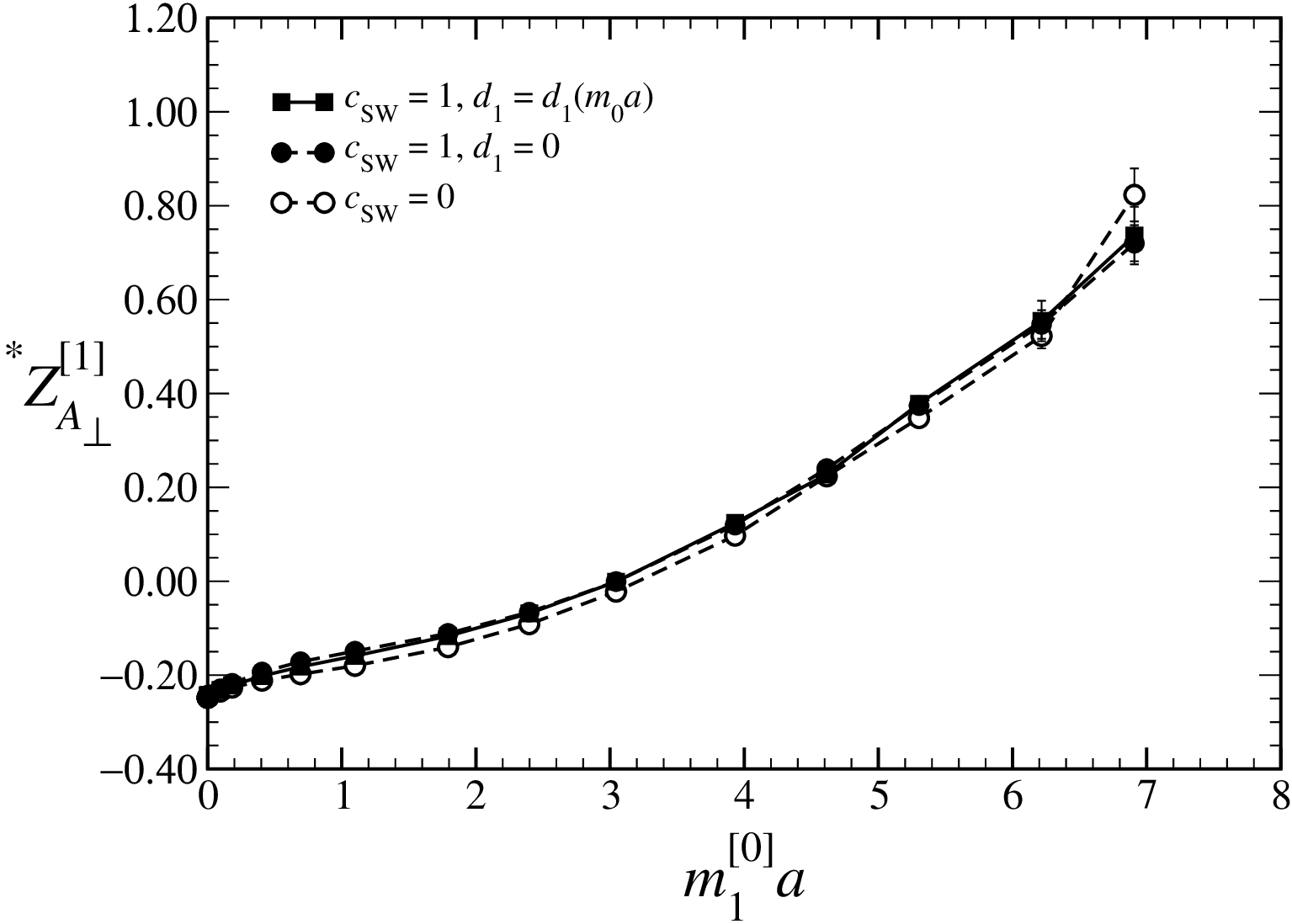} \\[0.5em]
	(c) \includegraphics[width=0.44\textwidth]{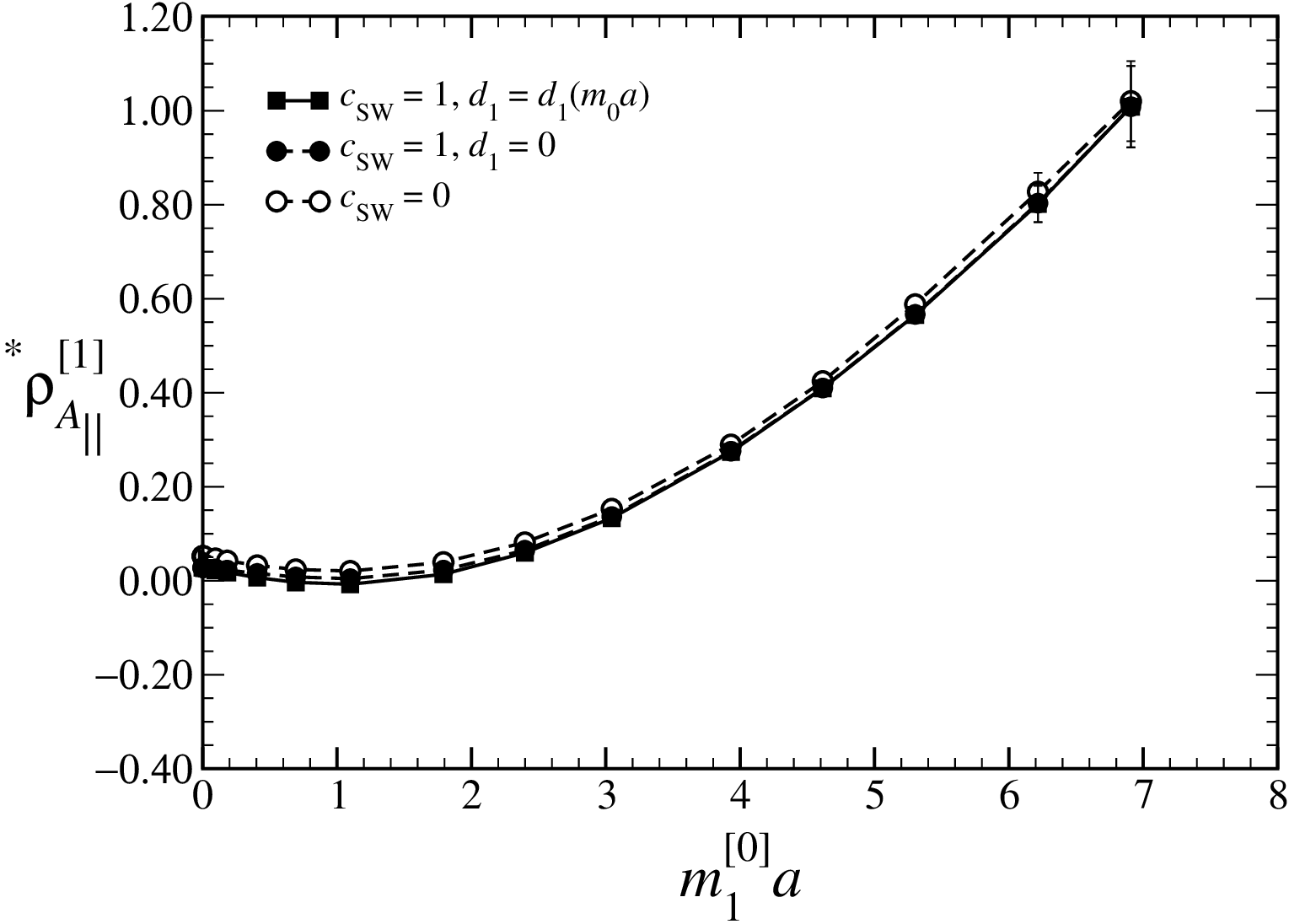} \hfill
	(d) \includegraphics[width=0.44\textwidth]{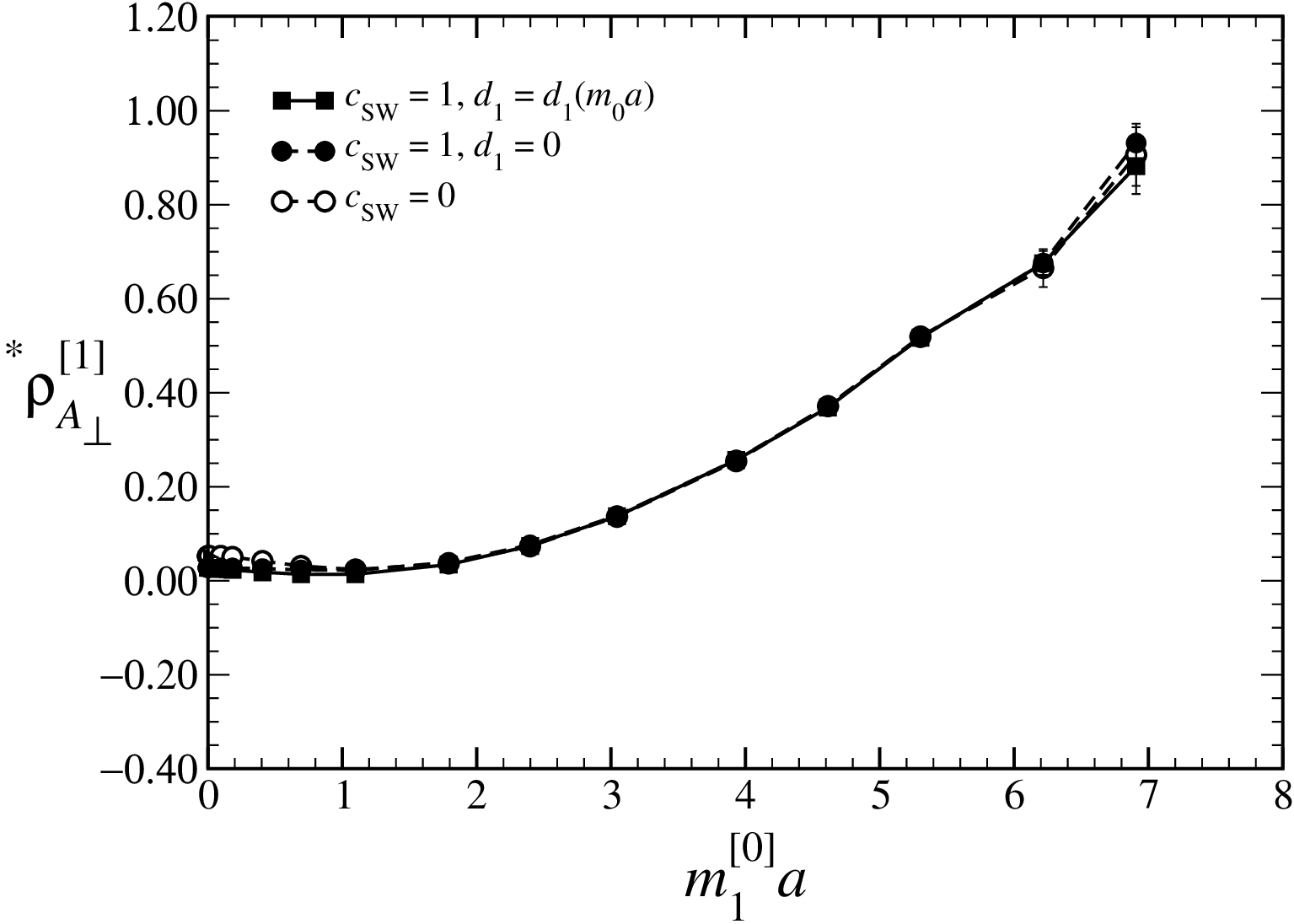}
	\caption[fig:starZA]{Full mass dependence of the estimated $\beta_0g^4$
	terms of the matching factors of the axial vector current
	   (a)~${}^*Z^{[1]}_{A_\parallel}$,        (b)~${}^*Z^{[1]}_{A_\perp}$,
	(c)~${}^*\rho^{[1]}_{A_\parallel}$, and (d)~${}^*\rho^{[1]}_{A_\perp}$.}
	\label{fig:starZA}
\end{figure}

Figure~\ref{fig:starZV} plots the full mass dependence of
the numerators for the vector current,
   (a)~${}^*Z^{[1]}_{V_\parallel}$,        (b)~${}^*Z^{[1]}_{V_\perp}$,
(c)~${}^*\rho^{[1]}_{V_\parallel}$, and (d)~${}^*\rho^{[1]}_{V_\perp}$.
As before, these numerical results are for the SW action with rotation
(solid lines) and also for the SW and Wilson actions without the
rotation (dotted lines).
Figure~\ref{fig:starZA} plots the full mass dependence of
the numerator of Eq.~(\ref{eq:qStar}) for the axial vector current,
   (a)~${}^*Z^{[1]}_{A_\parallel}$,        (b)~${}^*Z^{[1]}_{A_\perp}$,
(c)~${}^*\rho^{[1]}_{A_\parallel}$, and (d)~${}^*\rho^{[1]}_{A_\perp}$.
We have carried out several checks on our calculations.
Once again, identical numerical results have been obtained with two
or more completely independent programs.
Also, at $m_ba=0$ we reproduce the results, for the Wilson action,
of Ref.~\cite{Bernard:1999sx}.

For ${}^*Z_J^{[1]}$ and ${}^*\rho_J^{[1]}$ the limit of large $ma$
also has distinctive features.
In that case
\begin{equation}
	{}^*Z_J^{[1]} = \frac{C_F}{16\pi^2}\left[
		\case{1}{2} \gamma_h \ln^2(m_2a)^2 +
		\gamma'_h \ln(m_2a)^2 + {}^*z_J \right],
	\label{eq:starZinf}
\end{equation}
where $\gamma'_h$ is related to the two-loop anomalous dimension.
A~similar expression holds for ${}^*\rho_J^{[1]}$, with a different
constant.
Note that---in both cases---the one-loop anomalous dimension
appears multiplying~$\ln^2(m_2a)$.
The growth expected from Eq.~(\ref{eq:starZinf}) is seen in
Figs.~\ref{fig:starZV} and~\ref{fig:starZA}.
As a consequence, one finds $q^*a\propto\sqrt{m_2a}$ as $ma\to\infty$.
Square root behavior is typical of cases with an anomalous dimension.

\begin{figure}
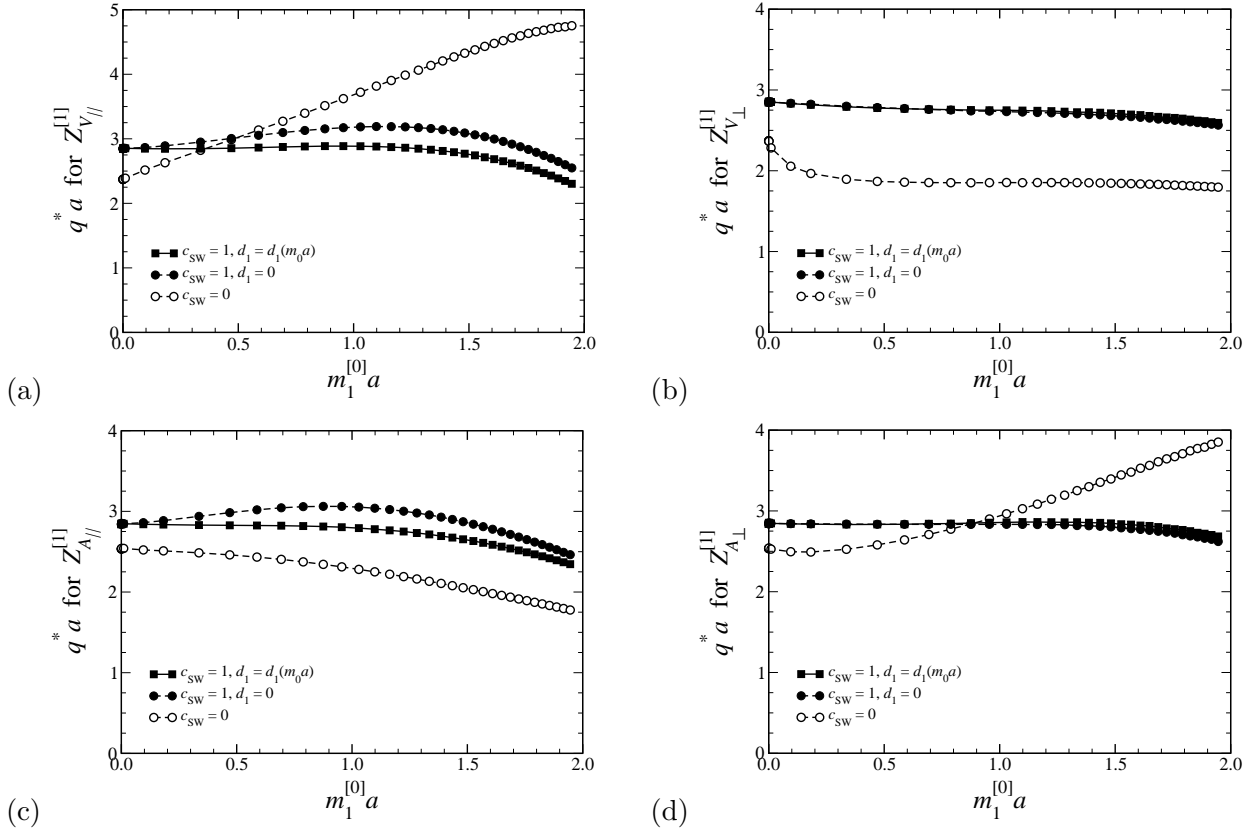

	\centering\small
	(a) \includegraphics[width=0.44\textwidth]{qstarZV4m1.eps} \hfill
	(b) \includegraphics[width=0.44\textwidth]{qstarZVjm1.eps} \\[0.5em]
	(c) \includegraphics[width=0.44\textwidth]{qstarZA4m1.eps} \hfill
	(d) \includegraphics[width=0.44\textwidth]{qstarZAjm1.eps}
    \caption[fig:qstar]{Full mass dependence of the BLM scale $q^*$, for
    (a)~$Z_{V_\parallel}$,     (b)~$Z_{V_\perp}$,
    (c)~$Z_{A_\parallel}$, and (d)~$Z_{A_\perp}$.}
    \label{fig:qstar}
\end{figure}

For the $Z$ factors, the resulting values for $q^*a$ are relatively
constant in the ``low mass'' region, $q^*a\sim 2.7$--2.9.
Figure~\ref{fig:qstar} shows how $q^*a$ depends on the heavy quark mass
in the region $m_1a\le 2$, which is the one most relevant to
calculations of decay constants and form factors.
At larger masses $Z^{[1]}$ goes through zero, at which point the original
BLM prescription breaks down.
A~prescription for~$q^*$ in this case is given in
Ref.~\cite{Hornbostel:2001ey}.
For the Wilson action the zero in $Z_{V_\parallel}^{[1]}$ is at a
smaller than usual mass [see Fig.~\ref{fig:ZV}(a)], which explains
its behavior for the BLM~$q^*a$ seen in Fig.~\ref{fig:qstar}(a).
For the $\rho$ factors the denominator $\rho^{[1]}$ is small over most
of the interesting region, as seen in Figs.~\ref{fig:ZV}(c)--(d)
and~\ref{fig:ZA}(c)--(d).

\begin{figure}
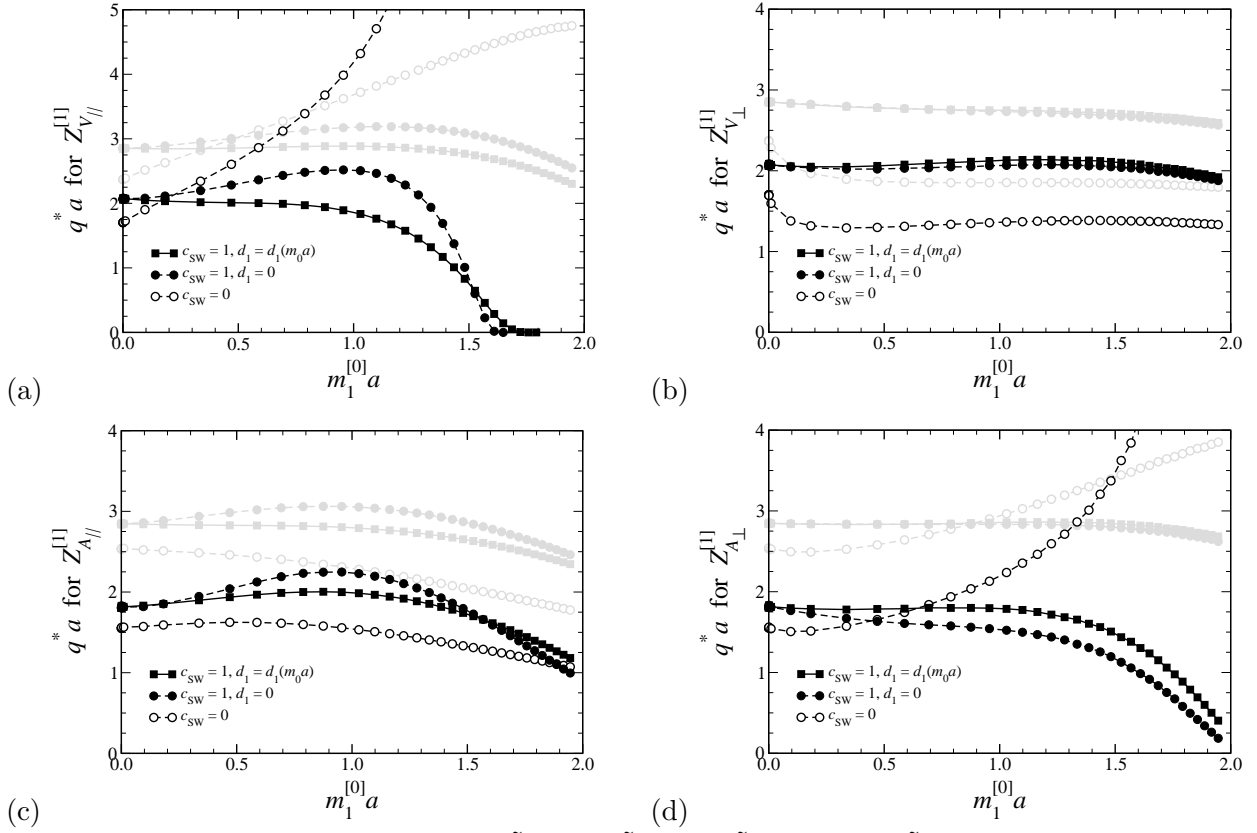

	\centering\small
	(a) \includegraphics[width=0.44\textwidth]{tadqstarZV4m1.eps} \hfill
	(b) \includegraphics[width=0.44\textwidth]{tadqstarZVjm1.eps} \\[0.5em]
	(c) \includegraphics[width=0.44\textwidth]{tadqstarZA4m1.eps} \hfill
	(d) \includegraphics[width=0.44\textwidth]{tadqstarZAjm1.eps}
    \caption[fig:tadqstar]{Tadpole-improved $q^*$ for
    (a)~$\tilde{Z}_{V_\parallel}$,     (b)~$\tilde{Z}_{V_\perp}$,
    (c)~$\tilde{Z}_{A_\parallel}$, and (d)~$\tilde{Z}_{A_\perp}$, with
	Fig.~\ref{fig:qstar} in gray.}
    \label{fig:tadqstar}
\end{figure}

It is also interesting to see how $q^*$ changes under tadpole
improvement.
If one introduces the tadpole-improved matching factors
\begin{equation}
	\tilde{Z}_J = Z_J/u_0,
\end{equation}
where the mean link $u_0$ is any tadpole-dominated short-distance
quantity, the arguments of Ref.~\cite{Lepage:1993xa} suggest that
the perturbative series for $\tilde{Z}_J$ has smaller coefficients.
In analogy with Eq.~(\ref{eq:ZJGamma}) we write
\begin{equation}
	e^{-\tilde{m}^{[0]}_1a/2} \tilde{Z}_J =
		1 + \sum_{l=1}^\infty g_0^{2l} \tilde{Z}_J^{[l]},
	\label{eq:tildeZJ[1]}
\end{equation}
where
\begin{equation}
	\tilde{m}_1^{[0]}a = \ln[1 + m_0a/u_0]
\end{equation}
is the tadpole-improved rest mass.
Then
\begin{equation}
	\tilde{Z}_J^{[1]} = Z_J^{[1]} - \frac{1}{2}
		\left(1 + \frac{1}{1+m_0a} \right) u_0^{[1]},
\end{equation}
and because $Z_J^{[1]}<0$ and $u_0^{[1]}<0$ one sees that the one-loop
coefficients are reduced.
Similarly, for computing the BLM scale
\begin{equation}
	{}^*\tilde{Z}_J^{[1]} = {}^*Z_J^{[1]} - \frac{1}{2}
		\left(1 + \frac{1}{1+m_0a} \right) {}^*u_0^{[1]}.
\end{equation}
To illustrate, we take $u_0$ from the average plaquette,
so $u_0^{[1]}=-C_F/16$ and $^*u_0^{[1]}=-0.204049(1)$.
Figure~\ref{fig:tadqstar} shows that, as a rule, $q^*$ is significantly
reduced, which means that tadpole improvement has removed some of the
most ultraviolet contributions.
With a lower scale, the coupling $g^2_V(q^*)$ becomes a bit larger
with tadpole improvement.
For $Z_{V_\parallel}$ and $Z_{A_\perp}$, however, the denominator
$\tilde{Z}_J^{[1]}$ already vanishes for $m_1a\approx1.5$--2.0, leading to
rapid growth in the BLM $q^*$ for the Wilson action,
and a zero in the BLM $q^*$ for the SW action.
One should again define $q^*$ in a more robust
way~\cite{Hornbostel:2001ey}.
Another choice for the mean field is $u_0=8\kappa_{\text{crit}}$.
It gives coefficients and BLM scales that lie between the unimproved
and tadpole-improved cases~\cite{Harada:2001xm}.

Our method also allows us to obtain the BLM scale for the improvement
coefficients in the {\sl Alpha} program.
Then we are in a position to compare BLM perturbation theory with
non-perturbative determinations of these coefficients.
We will give these results for $q^*$ and the mentioned comparison 
in another publication~\cite{Harada:2001zz}.

\section{Conclusions}
\label{sec:conclusions}

In this paper we have set up a matching procedure, based on~HQET,
for heavy-light currents.
It is valid for all $ma$, where $m$ is the heavy quark's mass and $a$ is
the lattice spacing, and to all orders in the gauge coupling.
It could be applied to lattice NRQCD, although here it is applied to
Wilson fermions.
In the latter case, HQET matching agrees with Symanzik matching
when $ma\ll1$.
In this way, HQET matching is a natural and attractive extension into
the regime $ma\not\ll1$, which is needed for heavy-quark phenomenology.

Our one-loop results for the SW action are of immediate value for
lattice calculations of $f_B$ and of form factors for the semi-leptonic
decay $B\to\pi l\nu$.
Indeed, our earlier one-loop results~\cite{Ishikawa:1997xh} (which
omitted the ``rotation'' terms in the current) were used for $f_B$
in Refs.~\cite{Aoki:1998ji,El-Khadra:1998hq,Bernard:1998xi,%
AliKhan:2000eg,Bernard:2000nv}, and our results were used for
semi-leptonic form factors in Ref.~\cite{El-Khadra:2001rv}.
In particular, we have obtained the BLM scale~$q^*$ for the
matching factors, which should reduce the uncertainty of one-loop
calculations.
Similarly, computing part of the normalization factor,
namely $\sqrt{Z_{V_\parallel^{qq}}Z_{V_\parallel^{bb}}}$,
non-perturbatively reduces the normalization uncertainty even
further~\cite{Hashimoto:2000yp,Simone:2000nv,El-Khadra:2001rv}.
(The heavy-heavy normalization factor $Z_{V_\parallel^{bb}}$ is defined
in our companion paper for heavy-heavy currents~\cite{Kronfeld:1999tk}.)

An outstanding problem at this time is the one-loop calculation of the
coefficients $B_{Ji}^{\text{lat}}$ of the dimension-four terms in the
HQET description.
A~calculation of these coefficients, and the subsequent
adjustment of the parameters~$b_{Ji}$ in the lattice currents,
would eliminate uncertainties of order~$\alpha_s\bar{\Lambda}/m$
and~$\alpha_s\bar{\Lambda}a$ in (future) calculations of heavy-quark
matrix elements.
The algebra quickly becomes voluminous, making this problem well-suited
to automated techniques~\cite{Nobes:2001pt}.

\acknowledgments
A.S.K. would like to thank Akira Ukawa and the Center for Computational
Physics for hospitality while part of this work was being carried out,
and the Aspen Center for Physics for a stimulating atmosphere while
part of the paper was being written.
S.H., A.S.K., and T.O. would also like to thank the Institute for 
Nuclear Theory at the University of Washington for hospitality while 
this paper was being finished.
S.H. and T.O. are supported by Grants-in-Aid of the Japanese Ministry of
Education (Nos.\ 11740162 and 12640279, respectively).
K.-I.I. and N.Y. are supported by JSPS Research Fellowships.
Fermilab is operated by Universities Research Association Inc.,
under contract with the U.S.\ Department of Energy.

\appendix

\section{Feynman Rules}
\label{app:feynman}

The needed propagators and vertices for quark-gluon interactions are
given in Ref.~\cite{Mertens:1998wx}.
Here we give the additional Feynman rules induced by the rotation term
of the heavy quark.
The additional rules are easy to derive by expressing the covariant
difference operator as~\cite{Kronfeld:1985zv}
\begin{equation}
 	D^\mu_{\text{lat}} = \left[T_{+\mu} - T_{-\mu} \right]/(2a),
\end{equation}
where
\begin{equation}
	T_{\pm\mu} = t_{\pm\mu/2}
		e^{\pm g_0aA_\mu} t_{\pm\mu/2},
\end{equation}
and $t_{\pm\mu/2}$ translates fields to its right by one-half
lattice spacing in the $\pm\mu$~direction.

There are three rules to give, with 0, 1, and 2 gluons emerging from the
vertex. 
Let the Dirac matrix of the current be~$\Gamma$.
Then,
\begin{eqnarray}
	\mbox{0-gluon} & = &
		\Gamma\left[1 + id_1\sum_r\gamma_r\sin(p_r)\right] ,
	\label{eq:0} \\
	\mbox{1-gluon} & = & g_0 t^a \; d_1 \,
		\Gamma \gamma_i \cos(p+\case{1}{2} k)_i ,
	\label{eq:1} \\
	\mbox{2-gluon} & = & i g_0^2 \case{1}{2}\{t^a, t^b\} \delta_{ij} \;
		d_1\, \Gamma \gamma_i \sin(p+\case{1}{2} k+\case{1}{2}\ell)_i ,
	\label{eq:2}
\end{eqnarray}
where momentum $p$ is quark momentum flowing into vertex, and
$k$ and $\ell$ are gluon momentum flowing into vertex.
As in Ref.~\cite{Mertens:1998wx}, the matrices $t^a$ are
anti-Hermitian, \emph{i.e.}, $U_\mu=\exp\left(g_0t^aA^a_\mu\right)$,
$\sum_{aj} t^a_{ij} t^a_{jk} = - C_F \delta_{ik}$,
and $\tr t^at^b=-\case{1}{2}\delta^{ab}$.

\section{Dirac Algebra}
\label{app:dirac}

To compute the vertex function, there are four diagrams to consider,
depicted in Fig.~\ref{fig:Feyn}:
\begin{figure}[!b]
	(a) \includegraphics[width=0.20\textwidth]{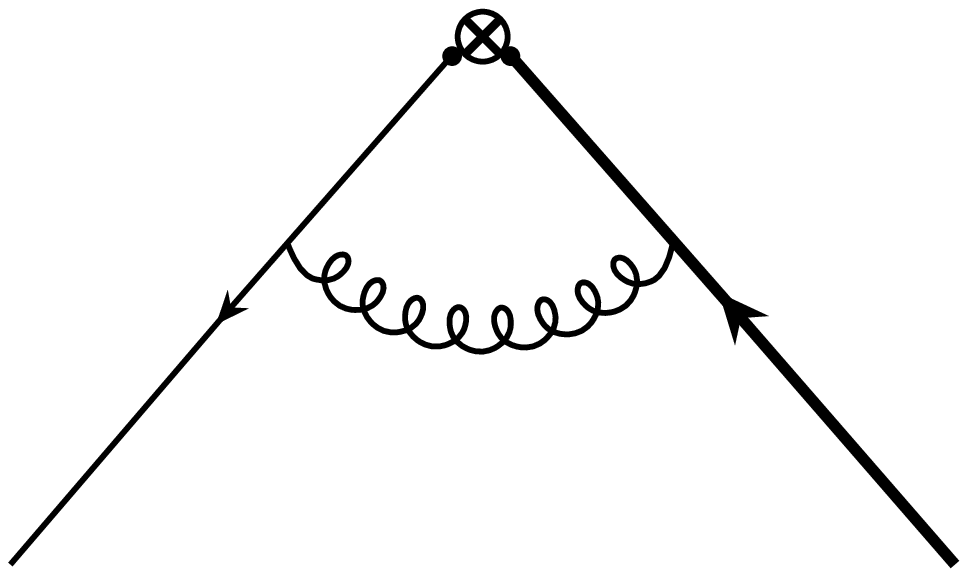} \hfill
	(b) \includegraphics[width=0.20\textwidth]{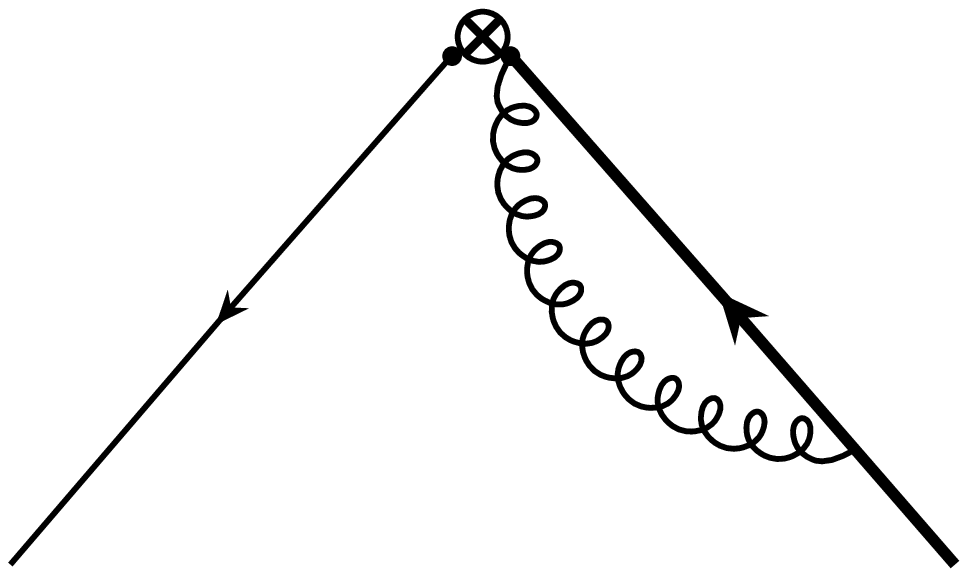} \hfill
	(c) \includegraphics[width=0.20\textwidth]{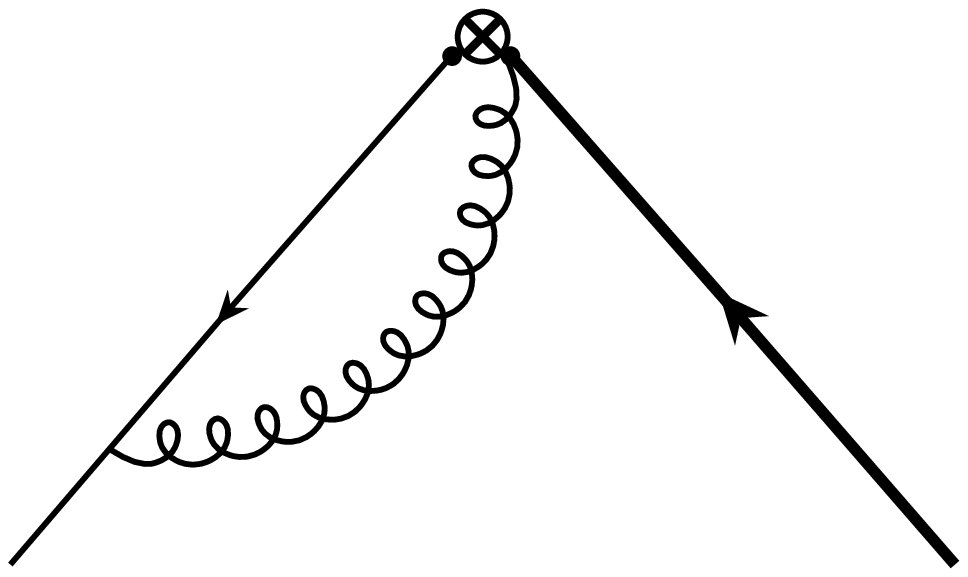} \hfill
	(d) \includegraphics[width=0.20\textwidth]{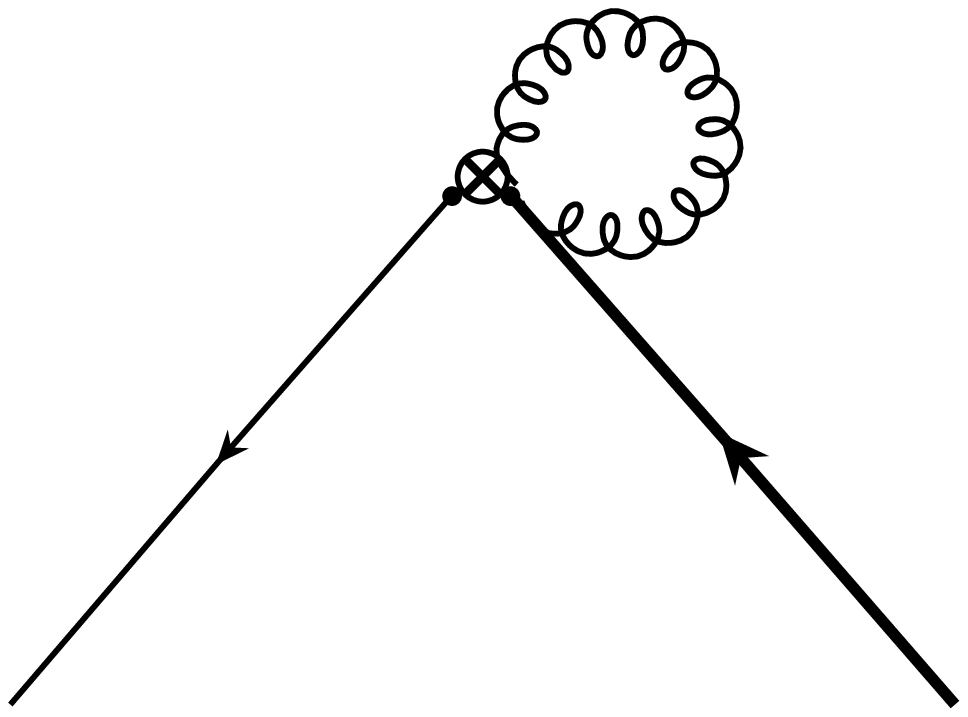}
	\caption[fig:Feyn]{Feynman diagrams for calculating the vertex
	function.
	The $\bullet$ on each side of the $\otimes$ indicates the rotation.}
	\label{fig:Feyn}
\end{figure}
the usual vertex diagram (with the rotation inside),
Fig.~\ref{fig:Feyn}(a);
two diagrams with the gluon connected to the incoming rotation,
Fig.~\ref{fig:Feyn}(b) and~(c);
and a tadpole diagram connected to the incoming rotation
[using rule~(\ref{eq:2})], Fig.~\ref{fig:Feyn}(d).
The tadpole diagram, Fig.~\ref{fig:Feyn}(d), vanishes for
zero external three-momentum, because $\ell=-k$ and $p_i=0$.

For each non-vanishing diagram, Figs.~\ref{fig:Feyn}(a--c),
define the integral
\begin{equation}
	I_\Gamma^{\text{(a,b,c)}} = - g_0^2 C_F \int \frac{d^4k}{(2\pi)^4}
		\frac{1}{\hat{k}^2}{\cal I}_\Gamma^{\text{(a,b,c)}},
\end{equation}
where $k$ is the momentum of the gluon in the loop, and
$\hat{k}_\mu=2\sin(\case{1}{2} k_\mu)$.
Let the incoming massive quark have couplings $m_0$, $r_s$, $\zeta$,
$c_B$, and $c_E$, and external momentum $p$.
Similarly, let the outgoing massless quark have couplings $m'_0=0$,
$r'_s$, $\zeta'$, $c'_B$, and $c'_E$, and external momentum~$p'$.
The internal quark lines carry momentum $p+k$ in and $p'+k$ out.
The integrals~$I$ are obtained directly from the loop diagrams.
Then
\begin{equation}
	Z_J^{[1]} =
		\frac{1}{2} \left(
		{Z_{2h}^{[1]}}_{\rm cont} - {Z_{2h}^{[1]}}_{\rm lat} +
		{Z_{2l}^{[1]}}_{\rm cont} - {Z_{2l}^{[1]}}_{\rm lat}
		\right) +
		\sum_{\rm d} \left(
		{I_\Gamma^{\rm d}}_{\rm cont} - {I_\Gamma^{\rm d}}_{\rm lat}
		\right),
\end{equation}
from Eq.~(\ref{eq:ZJGamma}).
The relation between the current $J$ and its Dirac matrix~$\Gamma$
is contained in Table~\ref{tab:s}.
\begin{table}
\centering
	\begin{tabular}{ccccc}
	&	   $J$   &      $\Gamma$      & $s_\Gamma$     & \\
		\hline
	&	  $V_4$  &     $\gamma_4$     & $-1$           & \\
	&	  $A_4$  & $\gamma_4\gamma_5$ & $+1$           & \\
	&	  $V_j$  &     $\gamma_j$     & $-\case{1}{3}$ & \\
	&	  $A_j$  & $\gamma_j\gamma_5$ & $+\case{1}{3}$ & \\
	\end{tabular}
	\caption[tab:s]{The factor $s_\Gamma$, defined by
		$\case{1}{3}\sum_r \gamma_r\Gamma\gamma_r=s_\Gamma \Gamma$.}
	\label{tab:s}
\end{table}
The expression relating ${Z_2^{[1]}}_{\rm lat}$ to lattice
self-energy functions is in Ref.~\cite{Mertens:1998wx}.

The most onerous task in evaluating the diagrams is the manipulation of
the Dirac matrices.
A convenient method is to treat each quark line separately, starting
from the initial- or final-state spinor.
Then the spinor, the propagator, and the vertices can be written out in
$2\times 2$ block diagonal form, with Pauli matrices appearing in the
blocks.
Once the Feynman rules are as complicated as in the present calculation,
it is easier to manipulate $2\times 2$ matrices of Pauli matrices than
to manipulate Dirac matrices.
A special advantage of this organization is that the rotation bracket in
Eq.~(\ref{eq:0}) merely ``rotates'' the rest of the leg.
We also obtain ${Z_2^{[1]}}_{\rm lat}$ in this way, with much less
effort than in Ref.~\cite{Mertens:1998wx}.

A further advantage is that the vertex corrections can be expressed
compactly.
The diagram with a gluon going from the incoming leg to the rotation,
Fig.~\ref{fig:Feyn}(b), is
\begin{equation}
	{\cal I}_\Gamma^{\text{(b)}} = d_1 \frac{\zeta}{D}
		\left[(3-\case{1}{4} \hat{\bbox{k}}^2) L
			+ \case{1}{2} \zeta \sum_r K_r S_r^2 \right],
\end{equation}
where $S_r=\sin k_r$, and the functions $D$, $L$, and $K_r$ are given in
Appendix~\ref{app:useful}.
The diagram with a gluon going from the outgoing leg to the rotation,
Fig.~\ref{fig:Feyn}(c), is
\begin{equation}
	{\cal I}_\Gamma^{\text{(c)}} = s_\Gamma d_1 \frac{\zeta'}{D'}
		\left[(3-\case{1}{4} \hat{\bbox{k}}^2) \maybebar{L}'
			+ \case{1}{2} \zeta' \sum_r K'_r S_r^2 \right],
\end{equation}
where the functions $D'$, $\maybebar{L}'$, and $K'_r$ are given in
Appendix~\ref{app:useful}, and~$s_\Gamma$ is given in Table~\ref{tab:s}.
The unbarred function~$L$ (barred function~$\bar{L}$) is for
$\Gamma=\gamma_4$ and $\gamma_j\gamma_5$
($\Gamma=\gamma_j$ and $\gamma_4\gamma_5$).

The vertex diagram, Fig.~\ref{fig:Feyn}(a), is complicated.
We find
${\cal I}_\Gamma^{\text{(a)}}=N_\Gamma^{\text{(a)}}/DD'$, with numerator
\begin{equation}
	N_\Gamma^{\text{(a)}} = 
	(\pm) \left( \maybebar{U}'_0 \R{U_0} -
	    s_\Gamma \maybebar{L}'_0 \R{L_0} \bbox{S}^2 \right)
		- \zeta\zeta' X_\Gamma.
\end{equation}
The upper sign and unbarred functions
         (lower sign and barred functions) are for
$\Gamma=\gamma_4$ and $\gamma_j\gamma_5$
($\Gamma=\gamma_j$ and $\gamma_4\gamma_5$).
The part $X_\Gamma$ comes from spatial gluon exchange:
\begin{eqnarray}
	X_\Gamma & = &
		- s_\Gamma (3-\case{1}{4} \hat{\bbox{k}}^2) \maybebar{L}' \R{L}
		+ s_\Gamma^2 (3-\case{1}{4} \hat{\bbox{k}}^2) \maybebar{V}' \R{V}
			\bbox{S}^2 \nonumber \\ & &
		+ \case{1}{2}   \left( \maybebar{V}'\R{U} - s_\Gamma \maybebar{L}'
		\R{\zeta}\right)
			\sum_r K_rS_r^2
				\nonumber \\ & &
		+ \case{1}{2}   \left( \maybebar{U}'\R{V} - s_\Gamma \zeta' \R{L}\right)
			\sum_r K'_rS_r^2
				\\ & &
		+ \case{1}{4}\left( \maybebar{U}'\R{U} - s_\Gamma \bbox{S}^2 \zeta'
		\R{\zeta}\right)
			\sum_r K'_rK_r\hat{k}_r^2
				\nonumber \\ & &
		+ \case{1}{8} (1-s_\Gamma^2)
		\left(\hat{\bbox{k}}^2\bbox{S}^2-3\sum_r\hat{k}_r^2S_r^2\right)
		\maybebar{V}' \R{V}, \nonumber
\end{eqnarray}
where the last term is absent for $V_4$ and $A_4$ (\emph{i.e.}, when
$s_\Gamma^2=1$).
The rotation enters in the ``rotated'' functions
\begin{eqnarray}
	\R{U_0} & = & U_0 + d_1 \bbox{S}^2 L_0, \\
	\R{L_0} & = & L_0 - d_1 U_0, \\
	\R{U} & = & U + d_1 \bbox{S}^2 \zeta, \\
	\R{\zeta} & = & \zeta - d_1 U, \\
	\R{V} & = & V + d_1 L, \\
	\R{L} & = & L - d_1 \bbox{S}^2 V.
\end{eqnarray}
Although the vertex diagram is not easy to write down, the rotation
modifies it in a fairly simple way, when using the $2\times 2$ Pauli
matrix method described above.

We have verified that these expressions are correct by completely
independent calculation with more common methods for the Dirac algebra.

\section{Useful Functions}
\label{app:useful}

In this appendix we list the functions appearing in
Appendix~\ref{app:dirac} for the action and currents given in
Sec.~\ref{sec:lattice}.
First, let
\begin{eqnarray}
	\mu  & = & 1 + m_0  + \case{1}{2} r_s  \zeta  \hat{\bbox{k}}^2 ,\\
	\mu' & = & 1 + m'_0 + \case{1}{2} r'_s \zeta' \hat{\bbox{k}}^2 .
\end{eqnarray}
from now on a prime means to replace incoming couplings and momenta
with corresponding outgoing couplings and momenta.

When the quark propagator is rationalized it has the denominator
\begin{equation}
	D = 1 - 2\mu\cos(k_4+im^{[0]}_1) + \mu^2 + \zeta^2 \bbox{S}^2,
\end{equation}
where $m_1^{[0]}=\ln(1+m_0)$.

In this calculation, the heavy quark has zero three-momentum, so its
spinor consists only of upper components.
Depending on the matrix $\Gamma$ the heavy quark couples either to the
upper or lower components of the light quark.
With the upper components the unbarred functions arise, and with the
lower components (of the light quark) the barred functions arise.

To express the useful functions compactly, it is convenient to
introduce first
\begin{eqnarray}
		U	& = & \mu - e^{-m_1^{[0]}+ik_4} , \\
	\bar{U}	& = & \mu - e^{+m_1^{[0]}-ik_4} ,
\end{eqnarray}
because these combinations appear in the other functions.
Then
\begin{eqnarray}
	U_0 & = & U e^{+m_1^{[0]} - ik_4/2} 
		- \case{1}{2} \zeta^2 c_E \cos(\case{1}{2} k_4) \bbox{S}^2 , \\
	L_0 & = & \zeta\left[ e^{+m_1^{[0]} - ik_4/2} 
		+ \case{1}{2}         c_E \cos(\case{1}{2} k_4)
			\bar{U} \right] , \\
	 V  & = & \zeta\left[1 + \case{i}{2} c_E\sin(k_4)\right]
		+ \case{1}{2} c_B U , \\
	 L  & = & - \bar{U} \left[1 + \case{i}{2}
	 c_E\sin(k_4)\right] + \case{1}{2} c_B \zeta \bbox{S}^2 , \\
	K_r & = & r_s - c_B \cos^2(\case{1}{2} k_r) =
	         (r_s - c_B) + \case{1}{4} c_B \hat{k}_r^2 ,
\end{eqnarray}
and
\begin{eqnarray}
\bar{U}_0 & = & \bar{U} e^{-m_1^{[0]} + ik_4/2} 
		- \case{1}{2} \zeta^2 c_E \cos(\case{1}{2} k_4) \bbox{S}^2 , \\
\bar{L}_0 & = & \zeta\left[ e^{-m_1^{[0]} + ik_4/2} 
		+ \case{1}{2}         c_E \cos(\case{1}{2} k_4)
			U \right] , \\
 \bar{V}  & = & \zeta\left[1 - \case{i}{2} c_E\sin(k_4)\right]
		+ \case{1}{2} c_B \bar{U} , \\
 \bar{L}  & = & - U \left[1 - \case{i}{2}
	 c_E\sin(k_4)\right] + \case{1}{2} c_B \zeta \bbox{S}^2 .
\end{eqnarray}
The barred functions are obtained from unbarred counterparts by
putting $k\to-k$ and $m^{[0]}_1\to-m^{[0]}_1$, so there is no need to
introduce $\bar{K}_r=K_r$.
In the present calculation the barred functions arise only for the
outgoing massless quark, for which $m'_0={m'_1}^{[0]}=0$.

\end{document}